\documentstyle[12pt,a4]{article}
	\newcommand{\vs}[1]{\vspace{#1 mm}}
	\newcommand{\hs}[1]{\hspace{#1 mm}}
	\newcommand{\non}{\nonumber}
	\newcommand{\be}{\begin{equation}}
	\newcommand{\ee}{\end{equation}}
	\newcommand{\bea}{\begin{eqnarray}}
	\newcommand{\eea}{\end{eqnarray}}
	\newcommand{\bpic}{\begin{picture}}
	\newcommand{\epic}{\end{picture}}
	\newcommand{\dm}{\partial_{\mu}}
	\newcommand{\dmm}{\partial\hs{0.25}^\mu}
	\newcommand{\dn}{\partial_\nu}
	\newcommand{\pa}{\partial}
	\newcommand{\ra}{\rightarrow}
	\newcommand{\lra}{\longrightarrow}
	\newcommand{\hi}{\phi_1}
	\newcommand{\go}{\phi_2}
	\newcommand{\ds}{\displaystyle}
	
	\newcommand{\de}{\delta}
	\newcommand{\D}{\Delta}
	\newcommand{\del}[1]{\delta_{#1}}
	\newcommand{\ep}[1]{\epsilon_{#1}}

	\newcommand{\vepj}{\varepsilon_j}
        \newcommand{\vepi}{\varepsilon_i}
	\newcommand{\etj}{\eta_{j}}
	\newcommand{\eti}{\eta_{i}}
	\newcommand{\G}{\Gamma}
	\newcommand{\gmn}{g^{\mu\nu}}
	\newcommand{\g}{g^2}
	\newcommand{\k}{k^2}
	\newcommand{\kmu}{k^{\mu}}
	\newcommand{\knu}{k^{\nu}}
	\newcommand{\cl}{\cal L}
	\newcommand{\lam}{\lambda}
	\newcommand{\vfi}{\varphi}
	\newcommand{\ro}{\rho}
	\newcommand{\om}{\omega}
	\newcommand{\si}{\sigma}
	\newcommand{\Si}{\Sigma}
	\newcommand{\ups}{\upsilon}
	\newcommand{\te}{\theta}
	
	\newcommand{\f}[2]{\frac{#1}{#2}}
	\newcommand{\fr}[1]{\frac{1}{#1}}
        \newcommand{\mf}{m_1^2}
        \newcommand{\ms}{m_2^2}
        \newcommand{\mt}{m_3^2}
        \newcommand{\mh}{m_H^2}
	\newcommand{\mg}{m_G^2}
	\newcommand{\ma}{m_A^2}
	
	\newcommand{\mpl}{m_+^2}
	\newcommand{\mmi}{m_-^2}
	\newcommand{\mb}{\mbox{}}

\begin{document}
\begin{titlepage}

        \begin{flushright}
                hep-th/9601176\\[2mm]
                CLNS 96/1396\\[2mm]
		\today\\[2mm]
		(revised)
	\end{flushright}
       
	\begin{center}
 	   {\large\bf Effective Action\vs{2}\\ 
               of\vs{2}\\
           Spontaneously Broken Gauge Theories\vs{12}\\}
           S.-H. Henry Tye\vs{1}\\
           Yan Vtorov-Karevsky\vs{4}\\
	   {\em Newman Laboratory of Nuclear Studies\\
		Cornell University, Ithaca, NY 14853}\vs{10}\\ 
	{\bf Abstract}\\
	\end{center}


        The effective action of a Higgs theory should be
gauge-invariant. However, the quantum and/or thermal contributions to 
the effective 
potential seem to be gauge-dependent, posing a problem
for its physical interpretation. In this paper, we identify the source
of the problem and argue that in 
a Higgs theory, perturbative contributions should be evaluated with 
the Higgs fields in the polar basis, not in the 
Cartesian basis. Formally, this observation can be made from the derivation 
of the Higgs theorem, which we provide. 
We show explicitly that, properly defined,
 the effective action for the Abelian Higgs theory is gauge invariant
to all orders in perturbation expansion when evaluated in the 
covariant gauge in the polar basis. In particular, the effective potential 
is gauge invariant. We also show the equivalence 
 between the calculations in the covariant gauge 
in the polar basis and the unitary gauge.
These points are illustrated explicitly with the one-loop calculations of
the effective action.
With a field redefinition, we obtain the physical effective potential. 
The $SU(2)$ non-Abelian case is also discussed.
 
\end{titlepage}


\section{Introduction}

        Effective action plays a very important role in quantum field
theory, especially in theories with spontaneous symmetry breakings.
For example, we have in mind the inflationary phase (at $T=0$) or the
electroweak phase transition ($T\neq 0$) in the early universe, where
space-time dependence plays a prominent role. In these and other
important physical applications, the knowledge of the effective action
(not just the effective potential) is crucial.

        The classical action must be corrected by quantum effects (and by
thermal effects at finite temperature). Many calculations of such
quantum/thermal effects have been carried out for the effective
action \cite{review}. The effective potential term inside the 
effective action has received the most attention. Since different 
gauge-fixings were used in the literature, resulting in different 
formulae for the effective potential term \cite{cowe}, 
it is not clear which are the appropriate formulae to use. 

        After the inclusion of quantum effects, the effective action of a 
Higgs theory must preserve the original global and local symmetries of the 
classical theory. The effective action must 
obey the original global symmetry of the theory so that the (would-be)
Goldstone bosons will remain massless after spontaneous symmetry breaking.
Gauge invariance must be preserved so that the Higgs mechanism can 
proceed. Using these criteria, we shall re-examine the effective action 
in various gauges. 

	In this paper, we shall make the following simple observation.
It is an elementary fact that a problem with rotational 
symmetry should be tackled using polar coordinates, not Cartesian 
coordinates. The Goldstone modes in Higgs theories should be massless,
even when one moves away from the extrema of the effective potential,
as is necessary for some physics applications. However, away from the 
extrema, the would-be Goldstone modes in the Cartesian basis
have non-zero masses. It is these masses that bring the gauge dependence
into the perturbative calculations of the effective action.
As we shall see, the would-be Goldstone bosons in the Cartesian basis
are mis-identified. 
On the other hand, the would-be Goldstone bosons in the polar basis are
always massless, at or away from the minimum, as they should be.
To correctly determine the effective action of a Higgs theory,
perturbation expansion should be carried out
in terms of appropriate polar variables. The unitary gauge,
and the covariant gauge with the Higgs field in the polar basis are thus
appropriate, but the usual covariant and $R_\xi$ gauges \cite{fuj} are not.

       Suppose the Higgs theory has certain global symmetry. 
It follows that the effective potential will also have that symmetry. 
After spontaneous symmetry breaking, there are (would-be) Goldstone bosons
corresponding to the generators of the broken symmetries. 
Some of them will be eaten by gauge bosons via the Higgs mechanism.
This will happen whenever spontaneous symmetry breaking occurs (i.e.,
non-zero vacuum expectation values develop), even if we move 
away from the minimum of the effective potential. So, for the Higgs 
mechanism to take place, the Goldstone 
bosons must remain massless, even away from the minimum 
of the effective potential. This is true in the polar basis, but not true
in the Cartesian basis. These points become more clear in the formal 
proof of the Higgs theorem, which is presented in Sec.\hs{0.8}2. Although
there are a number of proofs of the  Goldstone theorem \cite{gsw,jona},
to our knowledge, this is the first proof of the Higgs theorem. 

	The above argument can also be seen in perturbation theory.
First, we show that the $R_\xi$ gauge and its  
variations are poor choices to use in the effective 
action calculations because the $R_\xi$ gauge-fixing  
breaks the global symmetry of the theory. As a consequence, the effective 
potential and other terms in the effective action are no longer globally 
invariant. Furthermore, mixed propagators cannot be avoided
in any practical calculation, as illustrated explicitly by the one-loop 
correction to the effective potential. This removes the main motivation
for the $R_\xi$ gauge. Since the breaking of global symmetry is a
consequence of using the $R_\xi$ gauge, it can be avoided if we use another 
gauge which preserves this symmetry. 

        Next, let us consider the gauge invariance condition. 
The key issues are very similar in both the Abelian and the
non-Abelian Higgs theories at both zero and finite temperatures. To be 
specific, let us consider the Abelian Higgs theory (i.e., scalar QED) 
at zero temperature. The effective action is expected to 
have the following form:
\bea
\label{ea-gf}
\G\,[\,\phi, A_\mu ]  = \hs{100}\\[2mm]
= \int\! d\,^4x\left\{\,-\,V[\phi]\, +\,
\fr{2}\,Z_1[\phi]\,\dm\phi_i\,\dmm\phi_i
\,+ \,\fr{2}\,Z_2[\phi]\,{\ds \f{\phi_i \phi_j}{\phi^2}}
\,\dm\phi_i\,\dmm\phi_j\, -
 \right.\non\\
\left.\mb{} - Z_3[\phi]\,g\ep{ij}\phi_i\dm\phi_j A^\mu \,+ 
\fr{2} Z_4[\phi]\,\g\phi^2 A_\mu A^\mu
-\,\fr{4}\, Z_A[\phi]\,
F_{\mu\nu}F^{\mu\nu}\,+\,
\ldots \non \right\}
\eea
where the Higgs field is expressed in Cartesian variables $\phi_i$,
with $\phi^2=\phi_1^2+\phi_2^2$.
Not displayed are terms that involve higher powers of $A_\mu (x)\;$ 
and/or higher derivatives. As the $Z_2[\phi]$-term  
has two derivatives and is explicitly gauge-invariant,  
it must be, apriori, included in Eq.(\ref{ea-gf}). 
When gauge-fixing is necessary, we shall use
 the covariant gauge, which preserves the global symmetry of the theory. 
The gauge-fixed effective action $\G_{cov}[\,\phi, A_\mu ]$ is now defined 
as the sum of $\G[\,\phi, A_\mu ]$ given in Eq.(\ref{ea-gf}) and the 
gauge-fixing term
\bea
\label{gfix}
\G_{gf}[\,\phi, A_\mu] = \int\! d\,^4x \left\{- Z_{gf}[\phi]\,{\ds \fr{2\,\xi}}\,(\dm A^\mu)^2 + \ldots\,\right\}
\eea
Also, any ghost terms will be included in $\G\,[\,\phi, A_\mu ]$, although for
our purpose they will be absent since we do not introduce 
ghost background fields. It is this $\G[\phi,\,A_\mu]$, not
$\G_{cov}[\phi,\,A_\mu]$, that one should use to construct soliton solutions
(magnetic flux, monopole, domain walls etc), to study the evolution of
the scalar field in the inflationary epoch, or the nucleation during the
electroweak phase transition, etc. This point is clear at the tree-level
already.

        Let us consider the classical limit of $\G\,[\,\phi, A_\mu ]$ first. 
In this 
limit, all $Z_i[\phi]=1$, except for $Z_2[\phi]$, which is zero. 
The classical action is invariant under the gauge transformation,
with $\phi(x)=\phi_1(x)+i\phi_2(x)$,
\bea
\label{g-tr}
\phi(x) & \ra &  \phi(x)\,e\,^{i\,\te(x)}\\
A_\mu(x) & \ra & A_\mu (x)+ \fr{g}\, \dm \te\,(x)\non
\eea
We have argued that $\G\,[\,\phi, A_\mu ]$ must be (globally) 
$U(1)$-invariant and  gauge-invariant.
This means that all the $Z_i[\phi]$ and $V[\phi]$ are $U(1)$-invariant, and 
so they are also trivially gauge-invariant. 
Invariance of $\G\,[\,\phi, A_\mu ]$ under the gauge transformation
(\ref{g-tr}) then implies the following conditions: 

\be
        Z_1 [\phi] = Z_3[\phi] = Z_4 [\phi]   \label{ZZZ}
\ee
If the gauge symmetry condition (\ref{ZZZ}) is not satisfied, the 
would-be Goldstone boson cannot be gauged away; this would imply a breakdown 
of the Higgs mechanism. Any $\G\,[\,\phi, A_\mu ]$ that is invariant under 
the gauge transformation (\ref{g-tr}) will be referred to as actively 
gauge-invariant. If all the $Z_i[\phi]$ and $V[\phi]$ 
do not depend on the gauge parameter $\xi$, $\G\,[\,\phi, A_\mu ]$ 
will be referred to as passively gauge-invariant. $\G\,[\,\phi, A_\mu ]$
is gauge-invariant if it is both actively and passively gauge-invariant.

        Let us consider $\G\,[\,\phi, A_\mu ]$ in the covariant gauge.
Carrying out perturbation expansion in the Cartesian $(\hi , \go)$ basis 
(the usual approach), a number of 
observations can be made on the one-loop contribution to 
$\G\,[\,\phi, A_\mu ]$. After separating the gauge-fixing 
term (\ref{gfix}), we get $\G\,[\,\phi, A_\mu ]$  
as given in Eq.(\ref{ea-gf}) with the following properties: 

        (1) All the $Z_i[\phi]$ and $V[\phi]$ are globally 
$U(1)$-invariant, as expected. 

        (2) All the $Z_i[\phi]$ and $V[\phi]$ 
explicitly depend on the gauge parameter $\xi$.

        (3) The gauge symmetry condition (\ref{ZZZ}) is satisfied.
     
        (4) $Z_2[\phi]$ is not zero. 

Thus $\G\,[\,\phi, A_\mu ]$ obtained in the $(\hi , \go)$ basis is
actively but not passively gauge-invariant, that is, it is
explicitly $\xi$--dependent. What is the resolution?
It is well-known that the $\xi$--dependence drops out at the extrema of
$V[\phi]$. However, for physics that depends on the shape of the
effective potential, one should look at the extrema of the effective action
$\G\,[\,\phi, A_\mu ]$. Such physics must depend not only on $V[\phi]$,
but also on the derivative terms in $\G\,[\,\phi, A_\mu ]$.
What if the $\xi$-dependence in $V[\phi]$ is
always cancelled by the $\xi$-dependence coming from the derivative terms,
so that all physically measurable quantities are gauge invariant? Can this
be the case ? Knowing the $Z_i[\phi]$ and $V[\phi]$ to one-loop, we can
check this possibility explicitly. Unfortunately, and somewhat surprisingly,
the answer to this question is negative, except in the massless limit
(see Sec. 5 and 8).

	It is useful to find the source of the $\xi$-dependence in
$\G\,[\,\phi, A_\mu ]$. An examination of the $Z_i[\phi]$ and $V[\phi]$
shows that the $\xi$-dependence comes from the would-be
Goldstone boson mass $m_G$. At the minimum of the effective potential, 
$m_G$ vanishes and the $\xi$-dependence drops out. Away from the minimum, 
the non-vanishing $m_G$ brings in the $\xi$-dependence. 
We expect the Higgs mechanism to take place whenever there is a non-zero
vacuum expectation value. However, as explained in Sec.\hs{0.8}2, the 
non-vanishing $m_G$ invalidates the Higgs mechanism. 
This non-vanishing $m_G$ is an artifact of the Cartesian 
$(\hi, \go)$ basis. How to circumvent this problem?
The solution is suggested by the presence of $Z_2[\phi]$.
As $Z_2[\phi]$ is non-zero, the kinetic terms in 
$\G\,[\,\phi, A_\mu ]$ are no longer of the canonical form 
in the ($\hi , \go$) basis. 
However, in terms of the polar variables ($\ro, \chi$) which are 
related to the ($\hi, \go$) basis through  
$\phi (x) = \phi_1(x)+i\phi_2(x)=\ro (x) e^{i\chi (x)}$,
\be
\label{carpol}
{\ds \f{\phi_i \phi_j}{\phi^2}}\dm\phi_i\,\dmm\phi_j = 
(\dm\ro)^2\hs{5} {\rm and}\hs{5} 
(\del{ij} - {\ds \f{\phi_i\phi_j}{\phi^2}})\dm\phi_i\dmm\phi_j
= \ro^2(\dm\chi)^2.
\ee
Thus both scalar kinetic terms in $\G\,[\,\phi, A_\mu ]$ 
appear naturally in the polar basis.
As the would-be Goldstone boson is always massless in the polar basis, 
the $\xi$-dependence problem disappears, and the Higgs theorem always holds
(see Sec.\hs{0.8}\ref{section-higgs}). 
These observations clearly suggest that the polar variables should be 
used in the perturbative calculations of the effective action.

	When the one-loop contributions are calculated in the polar gauge 
(short for the covariant gauge in the polar basis), we find that 
$\G\,[\,\phi, A_\mu ]$ 
is both passively and actively gauge-invariant, {\em i.e.}  all the 
$Z_i[\phi]$ and $V[\phi]$ are $\xi$-independent and
Eq.(\ref{ZZZ}) holds. In fact, it is not difficult to show that 
$\G\,[\,\phi, A_\mu ]$ is gauge-invariant to all orders in the perturbation
expansion in this gauge, including higher derivative terms. 
 Thus the effective action calculated in the
polar basis satisfies the imposed criteria.
We may express $\G\,[\,\phi, A_\mu ]$ in the polar basis:
\bea
\label{earx}
\G[\, \phi, A_\mu] 
= \!\!\int\! d\,^4x\left\{- V[\ro] \,+\, 
\fr{2}\,Z_\ro [\ro]\,(\dm\ro)^2\, +\, \fr{2}\,
Z_1[\ro]\,\ro^2\,(\dm\chi -  g A_\mu)^2 \,- \right.\\[2mm]
\left.\mb -\,\fr{4}\,Z_A[\ro]\,
F_{\mu\nu}F^{\mu\nu}\!
+ \ldots \right\}\hs{50}\non
\eea
where Eq.(\ref{ZZZ}) is used and $Z_\ro [\ro]= Z_1[\ro] + Z_2[\ro]$.
In this basis, all $Z_i[\ro] = 1$
in the classical limit. With quantum corrections included,
$Z_\ro [\ro] \neq  Z_1[\ro]$. 

With the gauge invariance condition (\ref{ZZZ}) at hand, independent
calculations of $Z_1[\ro]$, $Z_3[\ro]$ and $Z_4[\ro]$ are clearly redundant.
To avoid this redundancy, a single source term 
for the combination $B_\mu = A_\mu - \fr{g} \dm\chi$ can be introduced 
in the generating functional, instead of separate sources for 
$A_\mu$ and $\chi$. Not surprisingly, this brings us to the unitary
gauge. In this gauge, both $B_\mu$ and $\ro$ are 
invariant under the gauge transformation (\ref{g-tr}),
so the source terms for $B_\mu$ and $\ro$ are also gauge invariant. 
Also, it is well-known  that 
only physical degrees of freedom have quantum fluctuations,
and no gauge-fixing is necessary in this case. So the 
resulting $\G[\phi, A_\mu]$ is by definition gauge-invariant.
It is easy to show that the contributions to the effective action in the
unitary gauge agree with the covariant gauge calculations in the polar
 basis to all orders in perturbation expansion. 
It should be noted that gauge boson propagators in the unitary gauge and
in the polar gauge are different (in particular their asymptotic behaviors).
 So agreement between the two calculations
provides a non-trivial check. In the polar gauge, we can see explicitly 
how the $\xi$-dependence drops out in $\G[\phi, A_\mu]$.
Unitary gauge calculations are more straightforward.

        As we have seen,  $Z_\ro [\ro]$ is non-trivial in general.
To define the physical effective potential, we introduce a field 
redefinition so that the resulting $Z_\ro$ is 1. In terms of this 
new Higgs field, the resulting effective potential should be
the physical effective potential that measures the 
energy density in the universe. As an illustration, we show how the 
Higgs mass is determined from the potential and compare the result to 
that obtained from standard perturbation theory.

        The plan of this paper is the following. In 
Sec.\hs{0.8}\ref{section-higgs}, we give 
a brief discussion of
both the Goldstone and the Higgs theorems. In Sec.\hs{0.8}\ref{section-prelim},
 we briefly review how to calculate the various $Z_i[\phi]$ and $V[\phi]$
in the effective action. 
In Sec.\hs{0.8}\ref{section-Rxi}, we argue why the $R_\xi$ 
gauge is a poor gauge choice 
for the determination of $\G[\phi,\,A_\mu]$. 
To illustrate this point clearly, we 
review, in Appendix A, the one-loop contribution to the effective potential 
in the $R_\xi$ gauge for the Abelian Higgs model.
In Sec.\hs{0.8}\ref{section-covariant}, we give the one-loop contribution 
to $\G[\phi,\,A_\mu]$ in the covariant gauge
in the ($\hi, \go$) basis (ignoring four and higher derivative terms). 
Details of the calculations are given in Appendices B and C. 
Here we pinpoint the source of the gauge dependence problem to the choice 
of the ($\hi, \go$) basis. We argue that the correct basis should be the 
polar one. In Sec.\hs{0.8}\ref{section-polar}, gauge-invariance of the 
effective action to all orders is shown in the polar gauge and some of the
one-loop contributions are provided.
Appendix D gives the Feynman rules in the polar and unitary gauges.
In Sec.\hs{0.8}\ref{section-unitary}, we briefly review and comment on
 the unitary gauge. Its equivalence to the polar gauge to all orders
is also shown.
In Sec.\hs{0.8}\ref{section-physEP}, we discuss the 
field redefinition needed to obtain a physical effective potential.
In Sec.\hs{0.8}\ref{section-nonabel}, we generalize our earlier discussion to
the $SU(2)$ non-Abelian Higgs theory. Sec.\hs{0.8}\ref{section-hot} 
contains a brief 
discussion of
the finite temperature case.
Sec.\hs{0.8}\ref{section-concl} contains concluding remarks.

\section{Higgs Theorem}
\label{section-higgs}
Here we present a brief discussion of the Goldstone and the Higgs theorems. 
Consider the 
generating functional $Z[J_a]$ of connected diagrams and the corresponding
effective action $\G\,[\psi_a]$:
\be
\G\,[\psi_a] = Z[J_a] - \int\hs{-1}d\,^4x\, J_a(x)\,\psi_a(x)\,,\hs{5}{\rm where}
\hs{5}
{\ds \f{\de\,\G}{\de \psi_a(x)}} = -\,J_a(x)  
\ee
The inverse propagator is given by	
\bea
D_{ab}
\hs{-2}\raisebox{1.5ex}
{$\!\scriptscriptstyle -1$}\, 
(x-y) = 
{\ds \f{\de^2\,\G}{\de\,\psi_a(x)\,\de\,\psi_b(y)}}\, =\,
-\, {\ds \f{\de\,J_a(x)}{\de\,\psi_b(y)}}
\eea

Now, variations of the sources can be expressed in terms of the inverse
propagators and variations of the fields,

\be
\de\,J_a(x) = -\!\int\hs{-1}d\,^4y\,D_{ab}
\hs{-3}\raisebox{1.5ex}
{$\scriptscriptstyle -1$}
(x-y)\,
\de\,\psi_b(y) 
\ee
Let us consider the Abelian Higgs theory in both the Cartesian and the 
polar bases. In the Cartesian basis, $\psi_a=(\phi_i,\,A_\mu )$. Under an 
infinitesimal gauge transformation $\te(x)$, $\de\,\psi_a(x)$ become
\bea
\de \hi = -\,\te\,\go\,,\hs{3}
\de \go = \te\,\hi\hs{5} {\rm and}\hs{5} \de A_\mu  =  
{\ds \fr{g}}\,\dm\te \non
\eea 
In the limit
$J_a \ra 0$, and $\de J_a \ra 0$, spontaneous symmetry breaking takes
place when $\phi$ developes a non-zero vacuum expectation value. 
We may take $\langle\hi\rangle = \ups$ and $\langle\go\rangle = 0$. 
In this case, the Fourier transforms of $\de\,J_2(x)$ and 
$\de\,J_\mu (x)$\, are given by
\bea
\label{22mu}
D_{22}\hs{-3}\raisebox{1.5ex}{$\scriptscriptstyle -1$}(k)\ups\,\te\,+\,
{\ds \fr{g}}\,D_{2\,\mu}\hs{-3}\raisebox{1.5ex}{$\scriptscriptstyle -1$}(k)
\,\kmu\te & = & 0\\
\label{2munu}
D_{\mu\,2}\hs{-3}\raisebox{1.5ex}{$\scriptscriptstyle -1$}(k)\ups\,\te \,+\,
{\ds \fr{g}}\,D_{\mu\nu}\hs{-3}\raisebox{1.5ex}{$\scriptscriptstyle -1$}(k)\,
\knu\te & = & 0
\eea
Note that for
the pure complex scalar theory, $\te$ is a non-zero constant and Eq.(\ref{22mu}) 
reduces to
\be
\label{22mu'}
D_{22}\hs{-3}\raisebox{1.5ex}{$\scriptscriptstyle -1$}(k)\,\ups\,\te = 0
\ee
At zero momentum, this is simply $m^2_2\ups\te=0$. When $\ups \neq 0$,
this equation implies that $\go$ describes a massless Goldstone boson.
This is the proof of the Goldstone theorem due to Jona-Lasinio \cite{jona}.
Writing $\G\,[\,\phi, A_\mu ]$ of the Abelian Higgs theory in the form 
given in Eq.(\ref{ea-gf}), Eqs.(\ref{22mu},\hs{0.5}\ref{2munu}) become, 
for small
momenta,
\bea
Z_1(\ups)\,(\k\hs{0.25}-\hs{0.25}m^2_2)\hs{0.25}
\ups\,\te \,- \,Z_3(\ups)\,\k\ups\,\te 
= 0\\
Z_3(\ups)\,\kmu g \ups^2\,\te \,- \,Z_4(\ups)\,{\ds \f{\ma}{g}}\,\kmu\,\te 
= 0
\eea
where terms with higher powers in momenta have been dropped.
Since $\G\,[\,\phi, A_\mu ]$ is gauge-invariant, {\em i.e.},
$Z_1(\ups) = Z_3(\ups) = Z_4(\ups)$, (which
follows from the gauge invariance condition (\ref{ZZZ})), 
we have $m_A = g\ups$ and $m_2 = 0$, as
expected. This means the gauge boson is massive when $\ups \neq 0$.
This is the Higgs theorem.
In this $(\hi,\go)$--basis, note that both Eq.(\ref{22mu}) and 
Eq.(\ref{22mu'}) are satisfied only at the minimum (or in general, any 
extremum) of the potential. This is because
\bea
\vs{-2}
m_2^2 = {\ds \fr{\ups}}{\ds \f{\pa V(\ups)}{\pa\ups}}
\vs{-2}\non\eea 
which is zero only at the extremum of the potential. 

If we now go to the polar basis, {\em i.e.} $\psi_a(x) = (\ro, \chi, A_\mu)$,
the infinitesimal gauge transformation becomes
\bea
\de \ro = 0\,,\hs{3}
\de \chi = \te\hs{5} {\rm and}\hs{5} \de A_\mu  =  
{\ds \fr{g}}\,\dm\te 
\eea
Following the same analysis, we have
\bea
\label{chichimu}
D_{\chi\chi}\hs{-3}\raisebox{1.5ex}{$\scriptscriptstyle -1$}(k)\te\,+\,
{\ds \f{i}{g}}\,D_{\chi\,\mu}\hs{-3}\raisebox{1.5ex}
{$\scriptscriptstyle -1$}(k)\,\kmu\te & = & 0\\
\label{chimunu}
i\hs{0.25}D_{\mu\,\chi}\hs{-3}\raisebox{1.5ex}{$\scriptscriptstyle -1$}(k)
\,\te \,-\,
{\ds \fr{g}}\,D_{\mu\nu}\hs{-3}\raisebox{1.5ex}{$\scriptscriptstyle -1$}(k)\,
\knu\te & = & 0
\eea
Since $\G\,[\ro, \chi, A_\mu]$ is explicitly $U(1)$--invariant, we have, for
any $\ups\neq 0$,
\bea
m_\chi^2=0,\hs{5}Z_1(\ups) = Z_3(\ups) = Z_4(\ups)  
\eea
For small momenta, we see that Eq.(\ref{chichimu}) 
is always satisfied, at or away from the minimum of the potential.
Eq.(\ref{chimunu}) gives $\ma = \g\ups^2$. To obtain the physical gauge 
boson mass, we must go to the pole of
the propagator (sitting at the minimum of the effective potential).

	Should we expect the Higgs theorem to hold when $\ups$ is not 
at the minimum of the potential? The answer is definitely {\em yes}.
Imagine one is measuring the energy density (or other physical
quantities) when passing through a magnetic flux in
a superconductor. Suppose one sits at a point 
where $\ups$ is not zero and not at the minimum of the potential. 
In this static situation, the energy density is physical and measurable,
and the photon is massive. This means the Higgs mechanism must be working,
i.e., there is a massless Goldstone boson that has been 
absorbed by the photon. This physics is correctly captured in the
polar basis but not in the Cartesian basis.

	In the pure complex scalar theory (i.e., turning off the
gauge coupling), the Goldstone theorem 
holds in the polar basis both 
at and away from the minimum of the potential. This is 
not the case in the Cartesian basis.
This implies that the proof of the Goldstone theorem should be 
carried out in the polar basis as well.
Although there are many versions of the proof of the Goldstone
theorem in the literature, to our knowledge, this is the first formal proof
of the Higgs theorem. Using the Lagrangian (\ref{non-ab}) which will be 
given in Sec.\ref{section-nonabel}, the generalization of the above argument 
to the
$SU(2)$ non-Abelian Higgs theory is straightforward.

\section{Preliminaries}
\label{section-prelim}
        Let us briefly review \cite{review} how to calculate  
quantum corrections
to the various $Z_j[\phi]$ and the effective potential $V[\phi]$ inside the
effective action $\G[\phi]$. Consider a set of the scalar fields $\phi_i$ 
in some 
theory. Suppose they fluctuate around some constants
$\ups_i$,
\be
\label{fluct}
\phi_i(x) = \ups_i + \vfi_i(x) \non
\ee
then the effective potential $V[\phi]$ can be Taylor-expanded around
$\ups_i$,
\bea
V[\phi(x)] = V(\ups) + \vfi_i(x)\,\pa_i V(\ups) + \fr{2}\vfi_i(x) \vfi_j(x) \,
\pa_i \pa_j V(\ups) + \ldots\\
{\rm where}\hs{10} \pa_i V(\ups) = 
{\ds \f{\pa V(\ups)}{\pa \ups_i}}\hs{65}\non 
\eea
Now we can 
determine $V[\phi]$ by calculating $V(\ups)$, which is given by the
one-particle irreducible (1PI)) pure-loop diagrams; alternatively, we can
calculate $\pa_i V(\ups)$, given by the 1PI tadpole diagrams, and perform 
an integration. One may also use $\pa_i \pa_j V(\ups)$,
given by the zero-momentum 1PI two-point functions, and then integrate twice
\ldots. This procedure is well-known. 
In particular, the 1-loop correction $V_1(\ups)$ to the 
effective potential can be calculated using the formula
\be
\label{lndet}
V_1(\ups) = - \f{i}{2} \int {\ds \f{d\hs{0.25}^4k}{(2\pi)^4}} 
\ln\det\left|iD^{-1}(\ups)\right|
\ee
where $iD^{-1}(\ups)$ is the quadratic part of the shifted Lagrangian.
Fermionic contributions carry an additional minus sign.
In this paper, dimensional regularization is used throughout.

        Analogously, one can expand any $Z_j[\phi]$ as

\be
\label{zexp}
Z[\phi(x)] = Z(\ups) + \vfi_i(x){\ds \f{\pa Z(\ups)}{\pa\ups_i}} + \ldots
\ee
For example, keeping the lowest term after this expansion,
the $Z_1[\phi]\,\dm\phi_i\,\dmm\phi_i$ term 
in $\G[\phi]$ becomes $Z_1(\ups)\,\dm\vfi_i\,\dmm\vfi_i$. 
Here $Z_1(\ups)$ 
can be calculated from the appropriate 1PI two-point 
diagrams, keeping only the terms that are quadratic in external momenta.
Of course, the actual perturbative calculations are sometimes 
complicated by the need for gauge-fixing. 

	If $V[\phi]$ has a particular global symmetry, the determination
of $V[\phi]$ simplifies substantially: all we need to do is
1) choose $\ups_1 \neq 0$ ,
$\ups_i = 0$, for $i = 2,...,n$, where these $n$ scalars form a
multiplet of the global symmetry;
2) calculate $V(\ups_1)$, the pure-loop diagrams, and
3) replace $\ups_1^2$ everywhere inside $V(\ups_1)$ by the globally 
symmetric $\phi^2(x)$ to obtain $V[\phi]$.

However, there is one essential subtlety about the calculations of $Z$'s.
There are, in general, two scalar field kinetic terms compatible with 
the global symmetry, and their coefficients
$Z_j[\phi]$ are apriori different. 
To properly separate $Z_1(\ups)$ from $Z_2(\ups)$ when these coefficients
are calculated from 1PI two-point diagrams, more than one
$\ups_i$ should be kept non-zero. 

\section{$R_\xi$ Gauge and Global Symmetry}
\label{section-Rxi}
	Suppose the Higgs theory has certain global symmetry that is gauged. 
The corresponding effective potential $V[\phi]$ must have the same global
symmetry as well. After spontaneous symmetry breaking with arbitrary $\ups$,
there are (would-be) Goldstone bosons
corresponding to the generators of the broken symmetries.
The Goldstone bosons must 
remain massless after the inclusion of quantum effects. 
Otherwise, the Higgs mechanism is ruined, as was pointed out 
 in Sec.\hs{0.8}2.
In light of this discussion, $R_\xi$ gauges are clearly unsuitable
for the effective action calculations,
since they break the global symmetry of the theory.
Let us discuss this in more details. We start with the Abelian Higgs 
Lagrangian (counterterms suppressed),
\bea
\label{aga}
{\cl}  =  -\fr{4}\,F_{\mu\nu}F^{\mu\nu}+\,\fr{2}\,
(D_\mu \phi)\raisebox{1ex}{\dag}(D^\mu \phi) - V[\phi]\hs{40}\\
{\rm where}\hs{10}D^\mu = \dmm - igA^\mu\,,\hs{10}V[\phi] = 
{\ds \f{m^2}{2}}\,\phi^2\,+\,{\ds \f{\lam}{4!}}\,\phi^4\hs{15}
\non\eea
Introducing the fluctuating fields via Eq.(\ref{fluct}),
we impose the $R_\xi$ gauge-fixing 
\bea
\label{rgft}
{\cl}_{gf}=\,{\ds -\,\fr{2\,\xi}}\,(\dm A^\mu -
\xi g\,\ep{ij}\,\vfi_{i}\,u_{j} )^2
\eea
where $u_1$ and $u_2$ are some constants. 
The 1-loop contribution
$V_{R_\xi,1}(\ups_i, u_i, \xi)$
to the effective potential was given in 
Ref.\cite{doja} and is reviewed in Appendix A.
Here we would like to make just a few remarks on that result:

	(a) Since $V_{R_\xi,\,1}(\ups_i, u_i, \xi)$ 
is not $U(1)$-invariant, we  
should keep both $\ups_1 \neq 0$ and $\ups_2 \neq 0$  in our calculations. 
If we set $\ups_2 = u_2 = 0$ (as done in \cite{fuku}) and calculate 
$V_{R_\xi,\,1}(\ups_1, 0, u_1, 0, \xi)$,
it will not determine the $\ups_2$ dependency in
$V_{R_\xi}(\ups_1, \ups_2, u_1, 0, \xi)$. To find this dependency,
one has to evaluate $\pa_2^2 V$, $\pa_2^4 V$  etc.,
-- the $2n$-point diagrams with external $\go$'s. These calculations 
are quite involved.

	(b) To obtain $V_{R_\xi}[\phi_i, u_i, \xi]$ from
 $V_{R_\xi}(\ups_i, u_i, \xi)$ , all 
$\ups_i$ are
 elevated to $\phi_i$, while $u_i$'s should be left untouched.
If one wants to avoid mixed propagators in the Feynman diagram
calculations (this was the original motivation for the $R_\xi$ gauge), one
sets $u_i = \ups_i$ (as done in \cite{wein}, where  $u_2 = \ups_2 = 0$). 
This prescription leaves one with an ambiguous 
$V_{R_\xi}(\ups_i, \xi)$ with some $\ups_i$'s
being part of the scalar field and thus elevated to $\phi_i$ and some not.
To find out which $\ups_i$ in $V_{R_\xi}(\ups_i, \xi)$ should be
elevated to $\phi_i$
 one must evaluate higher-point diagrams such as $\pa_i V$ etc.
 This calculation is clearly quite involved. In
practice, it is a lot easier to allow mixed propagators so one may determine
$V_{R_\xi}[\phi_i, u_i, \xi]$ via $V_{R_\xi}(\ups_i, u_i, \xi)$ alone. 

        (c) In the background field method \cite{abb}, 
the one-loop contribution still breaks the U(1)-invariance.

	(d) A way to remove the gauge-dependence in 
$V_{R_\xi}[\phi_i, u_i, \xi]$ was formally proposed in Ref.\cite{fuku,niel}.
This was explicitly implemented in the one-loop correction in 
Ref.\cite{aitch}. However, the result still depends explicitly on the
parameter $u_i$. Since the gauge-fixing introduces $u_i$, they should be
considered as gauge parameters. In this sense, $V_{R_\xi}[\phi_i, u_i]$ 
is still gauge-dependent. Even if one does not want to consider $u_i$ as 
gauge parameters, the resulting effective potential is not globally U(1)
invariant; so it is not actively gauge-invariant.

	(e) In the modified $R_\xi$ gauge of \cite{kas}, 
U(1)-invariance is again broken.
        
	(f) It is clear that if we set $u_i = 0$ (or $\xi = 0$) in 
Eq.(\ref{rgft}), the resulting $V_{R_\xi}(\ups_i)$ is $U(1)$ invariant. 
But then this is simply the covariant gauge result. In general,
covariant gauge obeys the global symmetry of the theory.

\section{Effective Action in Covariant Gauge}
\label{section-covariant}
        The effective action should maintain both the global and local
symmetries of the classical theory. After spontaneous symmetry breaking, 
we need this gauge symmetry to gauge away the (would-be) Goldstone bosons,
{\em i.e.}, the Higgs theorem is valid even at the quantum level (see 
Sec.\hs{0.8}2). 
 Also, perturbative quantum effects are not expected to destroy the
topological properties of gauge theories (e.g., the Aharonov-Bohm effect for
$U(1)$ gauge theory and instanton effects for non-Abelian gauge theories).
Since gauge invariance is crucial for the topological properties, the
effective action after quantum corrections must be gauge-invariant.

        Let us turn to the Abelian Higgs model in the covariant gauge. 
To carry out perturbative calculations, we must introduce a 
gauge-fixing term with the gauge parameter $\xi$. 
Since the 
effective action is expected to be globally $U(1)$-invariant, we can 
write the gauge-fixed effective action
$\G_{cov}[\phi, A_\mu]$ in the form 
\bea
\label{covea}
\G_{cov}[\phi, A_\mu] = \G[\phi, A_\mu] + \G_{gf}[\phi, A_\mu], 
\eea
where the effective action $\G[\phi, A_\mu]$ is given in Eq.(\ref{ea-gf})
 and the gauge-fixing term 
$\G_{gf}[\phi, A_\mu]$ is given in Eq.(\ref{gfix}).

	The $V[\phi,\xi]$ and $Z_i[\phi, \xi]$ will be calculated 
perturbatively.
The Feynman rules are well known and given in Appendix B for the sake 
of completeness. There is one point that we need to emphasize. To separate
$Z_1[\phi, \xi]$ and $Z_2[\phi, \xi]$, we need to keep both $\ups_1$ and
$\ups_2$ non-zero. This leads to mixed scalar propagators (which also 
mix with the gauge field propagator). As we shall see, the appearance of 
mixed scalar propagators is 
a clear signal that the $(\phi_1, \phi_2)$ basis is not the appropriate basis. 

	It is straightforward to evaluate the one-loop
contributions. Some of the relevant results are summarized in Appendix C. 
The effective potential in the covariant gauge can be obtained from 
that in the $R_\xi$-gauge by setting $u_i=0$, {\em i.e.},
$V[\phi,\xi]$=$V_{R_\xi}[\phi_i, 0, \xi]$. There are two types of 
derivative terms in $F_{\mu\nu}F^{\mu\nu}$,
one of which appears in the gauge-fixing term
$(\dm A^\mu)^2$. This allows us to uniquely separate $Z_A[\phi]$ and
$Z_{gf}[\phi]$ to obtain a gauge-invariant term in $\G\,[\,\phi, A_\mu ]$.
It is easy to see that $Z_{gf}[\phi]$ is non-trivial.

It is not surprising that  $V[\phi,\xi]$ and all the $Z_i[\phi, \xi]$ are 
explicitly $U(1)$-invariant. However, they are all $\xi$-dependent,
so $\G[\phi, A_\mu]$ is passively gauge-dependent. Here we want to point 
out the two properties claimed in the introduction (the explicit 
expressions are given in Appendix C):

        (1) $\G[\phi, A_\mu]$ is actively
gauge-invariant, {\em i.e.}, it is invariant under the gauge transformation
Eq.(\ref{g-tr}), {\em i.e.} Eq.(\ref{ZZZ}) is satisfied, 
$Z_1 [\phi, \xi] = Z_3[\phi, \xi] = Z_4 [\phi, \xi]$.

        (2) $Z_2[\phi, \xi]$ is non-zero; or equivalently, 
$Z_\ro[\phi, \xi]\neq Z_1[\phi, \xi]$. 
Note that $Z_2(\ups, \xi)$ is 
finite without regularization. Note also that $Z_2(\ups, \xi)$ has a pole
at the minimum of the classical effective potential.

        Is it possible that $\xi$-dependence of
$\G\,[\,\phi, A_\mu ]$ is harmless ? Physical quantities are determined
at the minima of $\G\,[\,\phi, A_\mu ]$. Away from the minima of
$V[\phi, \xi]$ this means physics must also depend on the $Z_i[\phi, \xi]$.
What if the $\xi$-dependence in $V[\phi, \xi]$ is
always cancelled by the $\xi$- dependence in $Z_i[\phi, \xi]$,
so that all physically measurable quantities are gauge invariant? Can this
be the case ? Knowing the $Z_i[\phi, \xi]$ and $V[\phi, \xi]$ to one-loop,
we can check this possibility explicitly.
Unfortunately, and somewhat surprisingly, the answer to this question is 
negative (except in the massless limit, see Sec. 8). An easy way to 
check this point is to consider the so-called 
physical effective potential. Suppose there exist new (dressed) 
fields $\phi '$ and  $A_\mu '$ as functionals of $\phi$ and $ A_\mu$, 
such that $\G\,[\,\phi ', A_\mu ']$ is gauge-invariant (in particular,
$\xi$-independent). Then one may consider $\G\,[\,\phi ', A_\mu ']$
to be satisfactory, and blame all the $\xi$-dependence in 
$\G\,[\,\phi, A_\mu ]$ on the bad choice of field variables.
As a check, we can introduce a 
new (dressed) field $\phi '$ as a functional of $\phi$,
so that the $\phi '$ kinetic term becomes canonical ({\em i.e.}, 
$Z_\ro[\phi '] = 1$). It turns out that $V[\phi ']$ is still
$\xi$-dependent (except in the massless limit, {\em i.e.} $m^2 = 0\,$; 
we shall come back to this case later). So we conclude 
that the covariant gauge in the ($\hi, \go)$ basis is unsuitable
for the determination of $\G\,[\,\phi, A_\mu ]$.

     The problem with the $(\hi,\go)$--basis stems from
the presence of fictitious degrees of freedom.
An examination of $Z_i[\phi,\,\xi]$ and 
$V[\phi,\,\xi]$ shows that $\xi$-dependence 
comes from the would-be Goldstone mass $m_G$.
As shown in Appendices B and C, two fictitious particles
 with 
$\xi$-dependent masses $m_{\pm}$ exist in the $(\phi_1, \phi_2)$ basis:
\be
\label{bad-masses}
 m_+^2 m_-^2 = \xi\ma\mg \hs{10}{\rm and}
\hs{10} m_+^2 + m_-^2 = \mg
\ee
where $\mg = m^2 + \lam\phi^2/6\hs{1}$ and $\hs{1}m_A = \g\phi^2$ is 
the gauge boson mass.
At the minimum (or, more generally, extremum) 
of the effective potential, $\mg$ vanishes and
the $\xi$-dependence drops out. Away from the minimum, the non-vanishing
$\mg$ brings in the masses $m_{\pm}$ which in turn
bring the $\xi$-dependence into the $Z_i[\phi,\,\xi]$ and 
$V[\phi,\,\xi]$. As explained in Sec.\hs{0.8}2, non-zero $\mg$ invalidates the
Higgs mechanism.
How can we avoid this problem?

The presence of $Z_2[\phi,\,\xi]$ 
suggests that the Lagrangian density of the Abelian Higgs theory 
should be written in the $(\ro,\,\chi)$ basis, where
\be
\label{carpol2}
\dm\chi = \ep{ij}\,{\ds \f{\phi_i\dm\phi_j}{\phi^2}}\hs{10}{\rm and}
\hs{10}\ro = |\phi|
\ee
The corresponding kinetic terms are given in Eq.(\ref{carpol}).
This further suggests that perturbative expansion should be carried out in the
polar basis. In this basis the $\ro$--mass is simply the Higgs mass $m_H$,
while the (would-be) Goldstone boson $\chi$ is {\em always} massless
even when we move away from the minimum of the potential.

 The (naive) Goldstone mass $\mg$ in
the $(\hi,\go)$--basis can be expressed in terms of polar
variables:
\be
\label{bad-mg}
\mg = \left.{\ds \f{\pa^2 V[\phi]}{\pa\go^2}}\right|\raisebox{-3.0ex}
{$\stackrel{\hi\,=\,\ups,}{\scriptstyle \go\,=\,0}$} \hs{2}=\hs{2} 
\left.{\ds \fr{\ro}}{\ds \f{\pa V[\ro]}{\pa\ro}}\right|\raisebox{-2.5ex}
{$\ro\!=\!\ups$}
\ee
{\em i.e.} $\mg$ is proportional to the first derivative of the effective 
potential. Away from the minimum of the potential, $\mg$ is non-zero and
the above problem  with 
$\xi$-dependent masses $m_{\pm}$ (\ref{bad-masses}) appears. 
Since $V[\ro]$ is $U(1)$--invariant 
at or away from the minimum of the potential, there should always be a 
massless mode, as is the case in the polar basis. This means that $\mg$
in Eq.(\ref{bad-mg}) is unphysical.
Since one has to move away from the minimum in order
to determine the effective action, the polar basis 
is the correct basis to use, not the $(\hi,\go)$
one. This argument also applies to theories with only global symmetries.

\section{Polar Gauge}
\label{section-polar}
The $\G\,[\,\phi, A_\mu ]$ calculated in the polar basis in perturbation 
expansion is completely gauge-invariant. First, we calculate explicitly 
the one-loop contributions to the various terms in Eq.(\ref{earx}) and
find them gauge-invariant. 
Then we prove the complete gauge-invariance of
the multi-loop and higher derivative contributions to the
$\G\,[\,\phi, A_\mu ]$.

Using Eq.(\ref{carpol}) and Eq.(\ref{carpol2}),
the Langrangian density of the Abelian Higgs theory in the 
polar gauge ({\em i.e.}, the polar basis in the covariant gauge) is 
\bea
\label{L-polar}
 \cl & = & \fr{2}\,(\dm\ro)^2 \,+\, \fr{2}\,\ro^2(\dm\chi)^2 \,-\, 
g\ro^2 A_\mu\dmm\chi \,+\, 
\,\fr{2}\,\g A^2 \ro^2 - \non\\
& & \mb - \,\fr{4}\,F_{\mu\nu}F^{\mu\nu} \,-\, V[\ro]\, -\, 
\fr{2\,\xi}\,(\dm A^\mu)^2 \,+\, \ro\,\bar{c}\,c\, +\, \bar{f}\dm\dmm f
\eea
Here the Faddeev-Popov (fermionic) ghosts $\bar{c}$ and $c$ come
from the Jacobian $ \f{\partial(\hi,\go)}{\partial(\ro,\,\chi)} $ 
and have no kinetic term.
The ghosts $\bar{f}$ and $f$ come from the covariant gauge-fixing.
The gauge-fixed effective action is
\bea
\label{pola}
\G_{cov}[\phi, A_\mu]\! 
& = & \!\int\! d\,^4x\left\{\,-\,V[\hs{0.25}\ro] + 
\fr{2}\,Z_\ro [\hs{0.25}\ro](\dm\ro)^2 \,+\, \fr{2}\,
Z_1[\hs{0.25}\ro]\,\ro^2\left(\dm\chi - 
 gA_\mu\right)^2 - \right.\non\\
& & \left. \mb - \fr{4}\,Z_A[\hs{0.25}\ro]\, 
F_{\mu\nu}F^{\mu\nu}  \,-\, \fr{2\,\xi}\,Z_{gf}[\hs{0.25}\ro]\,(\dm A^\mu)^2 +
\ldots \right\}
\eea
After the shift $\ro\ra\ro+\ups$, the inverse ($\chi,\,A$)--propagator becomes:
\bea
\label{d-ab}
iD^{-1}_{ab}(k)
= \left(
\begin{array}{r}
\ups^2\k  
 \hs{30}  i\,g\,\ups^2 \knu\hs{25}\\[2mm]
   -i\,g\,\ups^2 \kmu  \hs{10} 
- \gmn(\k-\ma) + (1 - {\ds \fr{\xi}})\,\kmu\knu 
\end{array}
\right)
\eea
where $a,b = \chi,\mu = \chi, 0, 1, 2, 3$.
The Higgs and ghost inverse propagators are trivial.
Feynman rules in this gauge are given in Appendix D.

Using Eq.(\ref{lndet}) it is easy to evaluate $V_1[\ro]$ :
\bea
\label{EP-polar}
V_1(\ups)& =& - \f{i}{2} \int \hs{-1}{\ds \f{d\hs{0.25}^4k}{(2\pi)^4}}
\left\{\ln\left[\,- {\ds \fr{\xi}}\,(\k - \ma)\hs{0.25}^3\,(\k - \mh)
\,k^4\ups^2\right] \,-\right.\non\\[2mm]
&&\left.\mb -\,2\ln\ups\,- \,2\ln(\k)\right\} 
\eea
where $\ma = \g\ups^2,\,\mh = m^2 + \fr{2}\lam\ups^2$.
Up to irrelevant constants, $V_1(\ups)$ is $\xi$-independent. The $k^4\ups^2$
factor from the matter fields is cancelled by the last two terms coming
from the ghosts, leaving behind only the physical degrees of freedom.
In contrast, in the $(\hi,\go)$-basis, $k^4 \,\ra\,(\k\,-\,m^2_+)
(\k\,-\,m^2_-)$ which was the source of all problems with that basis. 

So we obtain the effective potential:
\bea
\label{polact}
V[\ro]\,=\,  {\ds \f{m^2 \ro^2}{2} }\, + \,{\ds \f{\lam\,\ro^4}{4!}}\, +\,
{\ds \f{\hbar}{64\pi^2}}
\left\{ m_H^4(\,\ln{\ds \f{\mh}{\mu^2}} - \f{3}{2}) +
3\, m_A^4(\,\ln{\ds \f{\ma}{\mu^2}} -\f{5}{6}) \hs{-0.5}\right\}
\eea
where $\mu^2$ is the renormalization scale.

To evaluate $Z_\ro[\ro]$, let us first consider the 1PI two-point 
$\ro\,$-function $\Si(p)$, the $p^2$-term of which determines $Z_\ro(\ups)$.
Its one-loop contributions are
\bea
\label{si-ro}
\Si\hs{0.25}'(p)&=& \f{i}{2}\int \hs{-1}{\ds \f{d\hs{0.25}^4k}{(2\pi)^4}}
\left\{D_H(k)(-i\lam\ups)D_H(p+k)(-i\lam\ups)\, + \right.\\[2mm]
&&\left.\mb + \,D_{ab}(k)\hs{0.25}V_{bc}^{\,3}(k,-(p+k))\hs{0.25}
D_{cd}(p+k)\hs{0.25}V_{da}^{\,3}(p+k,-k)\right\}\non
\eea
where the prime denotes that $p_\mu$-independent diagrams are omitted.
$iD_H$ is the Higgs propagator and $iD_{ab}$ is the 
$(\chi,\,A)$--propagator ({\em i.e.} the inverse of $\,iD^{-1}_{ab}\,$ given
by Eq.(\ref{d-ab})). 
$V_{ab}^{\,3}\,$ are the appropriate 3-point vertices.
This equation can be represented by the following diagrams (solid lines
for the $\ro$-field, dashed lines for the $\chi$-field, and wavy lines
 for the $A$-field):
\bea
\label{picture}
\bpic(74,16)(-4,13)
\put(32,16){\circle{32}}
\put(16,16){\line(-1,0){16}}
\put(48,16){\line(1,0){16}}
\put(16,16){\circle*{3}}
\put(48,16){\circle*{3}}
\put(24,-18){$(a)$}
\epic
+
\bpic(74,16)(-4,13)
\put(15,16){\circle*{3}}
\put(49,16){\circle*{3}}
\put(15,16){\line(-1,0){16}}
\put(49,16){\line(1,0){16}}

\put(32,32){\oval(4,4)[t]}
\put(36,32){\oval(4,4)[bl]}
\put(28,32){\oval(4,4)[br]}
\multiput(36,30)(0.2,0.1){4}{\line(1,0){0.1}}
\multiput(28,30)(-0.2,0.1){4}{\line(1,0){0.1}}
\multiput(36.8,30.4)(0.1,0.1){16}{\line(1,0){0.1}}
\multiput(27.2,30.4)(-0.1,0.1){16}{\line(1,0){0.1}}
\multiput(38.4,32.0)(0.2,0.1){4}{\line(1,0){0.1}}
\multiput(25.6,32.0)(-0.2,0.1){4}{\line(1,0){0.1}}
\put(39.2,30.4){\oval(4,4)[tr]}
\put(24.8,30.4){\oval(4,4)[tl]}
\multiput(41.2,30.4)(-0.1,-0.2){12}{\line(0,1){0.1}}
\multiput(22.8,30.4)(0.1,-0.2){12}{\line(0,1){0.1}}
\put(42.0,28.0){\oval(4,4)[bl]}
\put(22.0,28.0){\oval(4,4)[br]}
\multiput(42.0,26.0)(0.2,0.1){15}{\line(1,0){0.1}}
\multiput(22.0,26.0)(-0.2,0.1){15}{\line(1,0){0.1}}
\put(45.0,25.5){\oval(4,4)[tr]}
\put(19.0,25.5){\oval(4,4)[tl]}
\multiput(47.0,25.5)(-0.1,-0.2){4}{\line(0,1){0.1}}
\multiput(17.0,25.5)(0.1,-0.2){4}{\line(0,1){0.1}}
\multiput(46.6,24.7)(-0.1,-0.1){15}{\line(0,1){0.1}}
\multiput(17.4,24.7)(0.1,-0.1){15}{\line(0,1){0.1}}
\multiput(45.1,23.2)(-0.1,-0.2){7}{\line(0,1){0.1}}
\multiput(18.9,23.2)(0.1,-0.2){7}{\line(0,1){0.1}}
\put(46.5,22.0){\oval(4,4)[bl]}
\put(17.5,22.0){\oval(4,4)[br]}
\multiput(46.5,19.9)(0.1,0){15}{\line(1,0){0.1}}
\multiput(17.5,19.9)(-0.1,0){15}{\line(1,0){0.1}}
\put(48.0,18.0){\oval(4,4)[r]}
\put(16.0,18.0){\oval(4,4)[l]}

\put(32,0){\oval(4,4)[b]}
\put(36,0){\oval(4,4)[tl]}
\put(28,0){\oval(4,4)[tr]}
\multiput(36,2)(0.2,-0.1){4}{\line(1,0){0.1}}
\multiput(28,2)(-0.2,-0.1){4}{\line(1,0){0.1}}
\multiput(36.8,1.6)(0.1,-0.1){16}{\line(1,0){0.1}}
\multiput(27.2,1.6)(-0.1,-0.1){16}{\line(1,0){0.1}}
\multiput(38.4,0.0)(0.2,-0.1){4}{\line(1,0){0.1}}
\multiput(25.6,0.0)(-0.2,-0.1){4}{\line(1,0){0.1}}
\put(39.2,1.6){\oval(4,4)[br]}
\put(24.8,1.6){\oval(4,4)[bl]}
\multiput(41.2,1.6)(-0.1,0.2){12}{\line(0,1){0.1}}
\multiput(22.8,1.6)(0.1,0.2){12}{\line(0,1){0.1}}
\put(42.0,4.0){\oval(4,4)[tl]}
\put(22.0,4.0){\oval(4,4)[tr]}
\multiput(42.0,6.0)(0.2,-0.1){15}{\line(1,0){0.1}}
\multiput(22.0,6.0)(-0.2,-0.1){15}{\line(1,0){0.1}}
\put(45.0,6.5){\oval(4,4)[br]}
\put(19.0,6.5){\oval(4,4)[bl]}
\multiput(47.0,6.5)(-0.1,0.2){4}{\line(0,1){0.1}}
\multiput(17.0,6.5)(0.1,0.2){4}{\line(0,1){0.1}}
\multiput(46.6,7.3)(-0.1,0.1){15}{\line(0,1){0.1}}
\multiput(17.4,7.3)(0.1,0.1){15}{\line(0,1){0.1}}
\multiput(45.1,8.8)(-0.1,0.2){7}{\line(0,1){0.1}}
\multiput(18.9,8.8)(0.1,0.2){7}{\line(0,1){0.1}}
\put(46.5,10.0){\oval(4,4)[tl]}
\put(17.5,10.0){\oval(4,4)[tr]}
\multiput(46.5,12.1)(0.1,0){15}{\line(1,0){0.1}}
\multiput(17.5,12.1)(-0.1,0){15}{\line(1,0){0.1}}
\put(48.0,14.0){\oval(4,4)[r]}
\put(16.0,14.0){\oval(4,4)[l]}

\put(24,-18){$(b)$}
\epic
+
\bpic(74,16)(-4,13)
\put(32,31.5){\line(-1,0){2}}
\put(16.5,16){\line(0,1){3}}
\multiput(17.4,22.3)(0.1,0.17){20}{\line(1,0){0.1}}
\multiput(22.3,28.6)(0.17,0.1){20}{\line(1,0){0.1}}
\put(32,31.5){\line(1,0){2}}
\put(47.5,16){\line(0,1){3}}
\multiput(46.6,22.3)(-0.1,0.17){20}{\line(1,0){0.1}}
\multiput(41.7,28.6)(-0.17,0.1){20}{\line(1,0){0.1}}
\put(32,0.5){\line(-1,0){2}}
\put(16.5,16){\line(0,-1){3}}
\multiput(17.4,9.7)(0.1,-0.17){20}{\line(1,0){0.1}}
\multiput(22.3,3.4)(0.17,-0.1){20}{\line(1,0){0.1}}
\put(32,0.5){\line(1,0){2}}
\put(47.5,16){\line(0,-1){3}}
\multiput(46.6,9.7)(-0.1,-0.17){20}{\line(1,0){0.1}}
\multiput(41.7,3.4)(-0.17,-0.1){20}{\line(1,0){0.1}}

\put(16,16){\circle*{3}}
\put(48,16){\circle*{3}}
\put(16,16){\line(-1,0){16}}
\put(48,16){\line(1,0){16}}
\put(24,-18){$(c)$}
\epic
+
\bpic(80,16)(-4,13)
\put(16,16){\circle*{3}}
\put(48,16){\circle*{3}}
\put(48,16){\line(1,0){16}}
\put(16,16){\line(-1,0){16}}
\put(32,0){\oval(4,4)[b]}
\put(36,0){\oval(4,4)[tl]}
\put(28,0){\oval(4,4)[tr]}
\multiput(36,2)(0.2,-0.1){4}{\line(1,0){0.1}}
\multiput(28,2)(-0.2,-0.1){4}{\line(1,0){0.1}}
\multiput(36.8,1.6)(0.1,-0.1){16}{\line(1,0){0.1}}
\multiput(27.2,1.6)(-0.1,-0.1){16}{\line(1,0){0.1}}
\multiput(38.4,0.0)(0.2,-0.1){4}{\line(1,0){0.1}}
\multiput(25.6,0.0)(-0.2,-0.1){4}{\line(1,0){0.1}}
\put(39.2,1.6){\oval(4,4)[br]}
\put(24.8,1.6){\oval(4,4)[bl]}
\multiput(41.2,1.6)(-0.1,0.2){12}{\line(0,1){0.1}}
\multiput(22.8,1.6)(0.1,0.2){12}{\line(0,1){0.1}}
\put(42.0,4.0){\oval(4,4)[tl]}
\put(22.0,4.0){\oval(4,4)[tr]}
\multiput(42.0,6.0)(0.2,-0.1){15}{\line(1,0){0.1}}
\multiput(22.0,6.0)(-0.2,-0.1){15}{\line(1,0){0.1}}
\put(45.0,6.5){\oval(4,4)[br]}
\put(19.0,6.5){\oval(4,4)[bl]}
\multiput(47.0,6.5)(-0.1,0.2){4}{\line(0,1){0.1}}
\multiput(17.0,6.5)(0.1,0.2){4}{\line(0,1){0.1}}
\multiput(46.6,7.3)(-0.1,0.1){15}{\line(0,1){0.1}}
\multiput(17.4,7.3)(0.1,0.1){15}{\line(0,1){0.1}}
\multiput(45.1,8.8)(-0.1,0.2){7}{\line(0,1){0.1}}
\multiput(18.9,8.8)(0.1,0.2){7}{\line(0,1){0.1}}
\put(46.5,10.0){\oval(4,4)[tl]}
\put(17.5,10.0){\oval(4,4)[tr]}
\multiput(46.5,12.1)(0.1,0){15}{\line(1,0){0.1}}
\multiput(17.5,12.1)(-0.1,0){15}{\line(1,0){0.1}}
\put(48.0,14.0){\oval(4,4)[r]}
\put(16.0,14.0){\oval(4,4)[l]}
\put(24,-18){$(d)$}
\put(32,31.5){\line(-1,0){2}}
\put(16.5,16){\line(0,1){3}}
\multiput(17.4,22.3)(0.1,0.17){20}{\line(1,0){0.1}}
\multiput(22.3,28.6)(0.17,0.1){20}{\line(1,0){0.1}}
\put(32,31.5){\line(1,0){2}}
\put(47.5,16){\line(0,1){3}}
\multiput(46.6,22.3)(-0.1,0.17){20}{\line(1,0){0.1}}
\multiput(41.7,28.6)(-0.17,0.1){20}{\line(1,0){0.1}}
\epic
\\
\bpic(74,60)(-4,13)
\put(16,16){\circle*{3}}
\put(48,16){\circle*{3}}
\put(48,16){\line(1,0){16}}
\put(16,16){\line(-1,0){16}}

\multiput(32,0)(0,-0.2){10}{\line(0,1){0.1}}
\put(32,0){\oval(4,4)[bl]}
\put(28,0){\oval(4,4)[tr]}
\multiput(28,2)(-0.2,-0.1){4}{\line(1,0){0.1}}
\multiput(27.2,1.6)(-0.1,-0.1){16}{\line(1,0){0.1}}
\multiput(25.6,0.0)(-0.2,-0.1){4}{\line(1,0){0.1}}
\put(24.8,1.6){\oval(4,4)[bl]}
\multiput(22.8,1.6)(0.1,0.2){12}{\line(0,1){0.1}}
\put(22.0,4.0){\oval(4,4)[tr]}
\multiput(22.0,6.0)(-0.2,-0.1){15}{\line(1,0){0.1}}
\put(19.0,6.5){\oval(4,4)[bl]}
\multiput(17.0,6.5)(0.1,0.2){4}{\line(0,1){0.1}}
\multiput(17.4,7.3)(0.1,0.1){15}{\line(0,1){0.1}}
\multiput(18.9,8.8)(0.1,0.2){7}{\line(0,1){0.1}}
\put(17.5,10.0){\oval(4,4)[tr]}
\multiput(17.5,12.1)(-0.1,0){15}{\line(1,0){0.1}}
\put(16.0,14.0){\oval(4,4)[l]}

\multiput(31.8,32)(0,0.2){12}{\line(0,1){0.1}}
\put(32,32){\oval(4,4)[tr]}
\put(36,32){\oval(4,4)[bl]}
\multiput(36,30)(0.2,0.1){4}{\line(1,0){0.1}}
\multiput(36.8,30.4)(0.1,0.1){16}{\line(1,0){0.1}}
\multiput(38.4,32.0)(0.2,0.1){4}{\line(1,0){0.1}}
\put(39.2,30.4){\oval(4,4)[tr]}
\multiput(41.2,30.4)(-0.1,-0.2){12}{\line(0,1){0.1}}
\put(42.0,28.0){\oval(4,4)[bl]}
\multiput(42.0,26.0)(0.2,0.1){15}{\line(1,0){0.1}}
\put(45.0,25.5){\oval(4,4)[tr]}
\multiput(47.0,25.5)(-0.1,-0.2){4}{\line(0,1){0.1}}
\multiput(46.6,24.7)(-0.1,-0.1){15}{\line(0,1){0.1}}
\multiput(45.1,23.2)(-0.1,-0.2){7}{\line(0,1){0.1}}
\put(46.5,22.0){\oval(4,4)[bl]}
\multiput(46.5,19.9)(0.1,0){15}{\line(1,0){0.1}}
\put(48.0,18.0){\oval(4,4)[r]}

\put(32,0.4){\line(1,0){3}}
\put(47.5,16){\line(0,-1){3}}
\multiput(46.6,9.7)(-0.1,-0.17){20}{\line(1,0){0.1}}
\multiput(41.7,3.4)(-0.17,-0.1){20}{\line(1,0){0.1}}
\put(31.8,31.7){\line(-1,0){3}}
\put(16.5,16){\line(0,1){3}}
\multiput(17.4,22.3)(0.1,0.17){20}{\line(1,0){0.1}}
\multiput(22.3,28.6)(0.17,0.1){20}{\line(1,0){0.1}}
\put(24,-18){$(e)$}
\epic
+
\bpic(74,60)(-4,13)
\put(16,16){\circle*{3}}
\put(49,16){\circle*{3}}
\put(49,16){\line(1,0){16}}
\put(16,16){\line(-1,0){16}}

\multiput(31.8,32)(0,0.2){10}{\line(0,1){0.1}}
\put(32,32){\oval(4,4)[tr]}
\put(36,32){\oval(4,4)[bl]}
\multiput(36,30)(0.2,0.1){4}{\line(1,0){0.1}}
\multiput(36.8,30.4)(0.1,0.1){16}{\line(1,0){0.1}}
\multiput(38.4,32.0)(0.2,0.1){4}{\line(1,0){0.1}}
\put(39.2,30.4){\oval(4,4)[tr]}
\multiput(41.2,30.4)(-0.1,-0.2){12}{\line(0,1){0.1}}
\put(42.0,28.0){\oval(4,4)[bl]}
\multiput(42.0,26.0)(0.2,0.1){15}{\line(1,0){0.1}}
\put(45.0,25.5){\oval(4,4)[tr]}
\multiput(47.0,25.5)(-0.1,-0.2){4}{\line(0,1){0.1}}
\multiput(46.6,24.7)(-0.1,-0.1){15}{\line(0,1){0.1}}
\multiput(45.1,23.2)(-0.1,-0.2){7}{\line(0,1){0.1}}
\put(46.5,22.0){\oval(4,4)[bl]}
\multiput(46.5,19.9)(0.1,0){15}{\line(1,0){0.1}}
\put(48.0,18.0){\oval(4,4)[r]}

\multiput(31.8,0.2)(0,-0.2){10}{\line(0,1){0.1}}
\put(32,0){\oval(4,4)[br]}
\put(36,0){\oval(4,4)[tl]}
\multiput(36,2)(0.2,-0.1){4}{\line(1,0){0.1}}
\multiput(36.8,1.6)(0.1,-0.1){16}{\line(1,0){0.1}}
\multiput(38.4,0.0)(0.2,-0.1){4}{\line(1,0){0.1}}
\put(39.2,1.6){\oval(4,4)[br]}
\multiput(41.2,1.6)(-0.1,0.2){12}{\line(0,1){0.1}}
\put(42.0,4.0){\oval(4,4)[tl]}
\multiput(42.0,6.0)(0.2,-0.1){15}{\line(1,0){0.1}}
\put(45.0,6.5){\oval(4,4)[br]}
\multiput(47.0,6.5)(-0.1,0.2){4}{\line(0,1){0.1}}
\multiput(46.6,7.3)(-0.1,0.1){15}{\line(0,1){0.1}}
\multiput(45.1,8.8)(-0.1,0.2){7}{\line(0,1){0.1}}
\put(46.5,10.0){\oval(4,4)[tl]}
\multiput(46.5,12.1)(0.1,0){15}{\line(1,0){0.1}}
\put(48.0,14.0){\oval(4,4)[r]}

\put(31.8,31.7){\line(-1,0){3}}
\put(16.5,16){\line(0,1){3}}
\multiput(17.4,22.3)(0.1,0.17){20}{\line(1,0){0.1}}
\multiput(22.3,28.6)(0.17,0.1){20}{\line(1,0){0.1}}
\put(31.8,0.3){\line(-1,0){3}}
\put(16.5,16){\line(0,-1){3}}
\multiput(17.4,9.7)(0.1,-0.17){20}{\line(1,0){0.1}}
\multiput(22.3,3.4)(0.17,-0.1){20}{\line(1,0){0.1}}

\put(24,-18){$(f)$}
\epic
+
\bpic(74,60)(-4,13)
\put(16,16){\circle*{3}}
\put(48,16){\circle*{3}}
\put(16,16){\line(-1,0){16}}
\put(48,16){\line(1,0){16}}
\multiput(31.8,32)(0,0.2){10}{\line(0,1){0.1}}
\put(32,32){\oval(4,4)[tr]}
\put(36,32){\oval(4,4)[bl]}
\multiput(36,30)(0.2,0.1){4}{\line(1,0){0.1}}
\multiput(36.8,30.4)(0.1,0.1){16}{\line(1,0){0.1}}
\multiput(38.4,32.0)(0.2,0.1){4}{\line(1,0){0.1}}
\put(39.2,30.4){\oval(4,4)[tr]}
\multiput(41.2,30.4)(-0.1,-0.2){12}{\line(0,1){0.1}}
\put(42.0,28.0){\oval(4,4)[bl]}
\multiput(42.0,26.0)(0.2,0.1){15}{\line(1,0){0.1}}
\put(45.0,25.5){\oval(4,4)[tr]}
\multiput(47.0,25.5)(-0.1,-0.2){4}{\line(0,1){0.1}}
\multiput(46.6,24.7)(-0.1,-0.1){15}{\line(0,1){0.1}}
\multiput(45.1,23.2)(-0.1,-0.2){7}{\line(0,1){0.1}}
\put(46.5,22.0){\oval(4,4)[bl]}
\multiput(46.5,19.9)(0.1,0){15}{\line(1,0){0.1}}
\put(48.0,18.0){\oval(4,4)[r]}

\put(31.8,31.7){\line(-1,0){3}}
\put(16.5,16){\line(0,1){3}}
\multiput(17.4,22.3)(0.1,0.17){20}{\line(1,0){0.1}}
\multiput(22.3,28.6)(0.17,0.1){20}{\line(1,0){0.1}}
\put(32,0.5){\line(-1,0){2}}
\put(16.5,16){\line(0,-1){3}}
\multiput(17.4,9.7)(0.1,-0.17){20}{\line(1,0){0.1}}
\multiput(22.3,3.4)(0.17,-0.1){20}{\line(1,0){0.1}}
\put(32,0.5){\line(1,0){2}}
\put(47.5,16){\line(0,-1){3}}
\multiput(46.6,9.7)(-0.1,-0.17){20}{\line(1,0){0.1}}
\multiput(41.7,3.4)(-0.17,-0.1){20}{\line(1,0){0.1}}
\put(24,-18){$(g)$}
\epic
+
\bpic(80,60)(-4,13)
\put(16,16){\circle*{3}}
\put(49,16){\circle*{3}}
\put(16,16){\line(-1,0){16}}
\put(49,16){\line(1,0){16}}

\multiput(31.8,32)(0,0.2){10}{\line(0,1){0.1}}
\put(32,32){\oval(4,4)[tr]}
\put(36,32){\oval(4,4)[bl]}
\multiput(36,30)(0.2,0.1){4}{\line(1,0){0.1}}
\multiput(36.8,30.4)(0.1,0.1){16}{\line(1,0){0.1}}
\multiput(38.4,32.0)(0.2,0.1){4}{\line(1,0){0.1}}
\put(39.2,30.4){\oval(4,4)[tr]}
\multiput(41.2,30.4)(-0.1,-0.2){12}{\line(0,1){0.1}}
\put(42.0,28.0){\oval(4,4)[bl]}
\multiput(42.0,26.0)(0.2,0.1){15}{\line(1,0){0.1}}
\put(45.0,25.5){\oval(4,4)[tr]}
\multiput(47.0,25.5)(-0.1,-0.2){4}{\line(0,1){0.1}}
\multiput(46.6,24.7)(-0.1,-0.1){15}{\line(0,1){0.1}}
\multiput(45.1,23.2)(-0.1,-0.2){7}{\line(0,1){0.1}}
\put(46.5,22.0){\oval(4,4)[bl]}
\multiput(46.5,19.9)(0.1,0){15}{\line(1,0){0.1}}
\put(48.0,18.0){\oval(4,4)[r]}
\put(32,0){\oval(4,4)[b]}
\put(36,0){\oval(4,4)[tl]}
\put(28,0){\oval(4,4)[tr]}
\multiput(36,2)(0.2,-0.1){4}{\line(1,0){0.1}}
\multiput(28,2)(-0.2,-0.1){4}{\line(1,0){0.1}}
\multiput(36.8,1.6)(0.1,-0.1){16}{\line(1,0){0.1}}
\multiput(27.2,1.6)(-0.1,-0.1){16}{\line(1,0){0.1}}
\multiput(38.4,0.0)(0.2,-0.1){4}{\line(1,0){0.1}}
\multiput(25.6,0.0)(-0.2,-0.1){4}{\line(1,0){0.1}}
\put(39.2,1.6){\oval(4,4)[br]}
\put(24.8,1.6){\oval(4,4)[bl]}
\multiput(41.2,1.6)(-0.1,0.2){12}{\line(0,1){0.1}}
\multiput(22.8,1.6)(0.1,0.2){12}{\line(0,1){0.1}}
\put(42.0,4.0){\oval(4,4)[tl]}
\put(22.0,4.0){\oval(4,4)[tr]}
\multiput(42.0,6.0)(0.2,-0.1){15}{\line(1,0){0.1}}
\multiput(22.0,6.0)(-0.2,-0.1){15}{\line(1,0){0.1}}
\put(45.0,6.5){\oval(4,4)[br]}
\put(19.0,6.5){\oval(4,4)[bl]}
\multiput(47.0,6.5)(-0.1,0.2){4}{\line(0,1){0.1}}
\multiput(17.0,6.5)(0.1,0.2){4}{\line(0,1){0.1}}
\multiput(46.6,7.3)(-0.1,0.1){15}{\line(0,1){0.1}}
\multiput(17.4,7.3)(0.1,0.1){15}{\line(0,1){0.1}}
\multiput(45.1,8.8)(-0.1,0.2){7}{\line(0,1){0.1}}
\multiput(18.9,8.8)(0.1,0.2){7}{\line(0,1){0.1}}
\put(46.5,10.0){\oval(4,4)[tl]}
\put(17.5,10.0){\oval(4,4)[tr]}
\multiput(46.5,12.1)(0.1,0){15}{\line(1,0){0.1}}
\multiput(17.5,12.1)(-0.1,0){15}{\line(1,0){0.1}}
\put(48.0,14.0){\oval(4,4)[r]}
\put(16.0,14.0){\oval(4,4)[l]}
\put(31.8,31.7){\line(-1,0){3}}
\put(16.5,16){\line(0,1){3}}
\multiput(17.4,22.3)(0.1,0.17){20}{\line(1,0){0.1}}
\multiput(22.3,28.6)(0.17,0.1){20}{\line(1,0){0.1}}

\put(24,-18){$(h)$}
\epic\non\\[4mm]
\non
\eea  
where proper mirror reflections of the above diagrams 
must be included.
Diagram $(a)$ corresponds to the first term of Eq.(\ref{si-ro}), 
diagrams $(b-h)$
correspond to the second term.
Individual diagrams $(b-h)$ are $\xi$-dependent. 
To see that Eq.(\ref{si-ro}) is $\xi$-independent, let us note first that 
the $\xi$-dependence in Eq.(\ref{si-ro})
appears only in the $(\chi,\,A)$--propagator $iD_{ab}$. 
We may write $D_{ab}(k)$ as
\be
\label{polar-prop}
D_{ab}(k) = D_{ab}^{(0)}(k) \,+\, \xi\hs{0.25} D_{ab}^{(1)}(k)
\ee
where 
\bea
\label{d0-ab}
D_{ab}^{(0)}(k) = 
\left(
\begin{array}{r}
\hs{-2}{\ds \fr{\k\ups^2}}  
 \hs{22}0\hs{17}  \\[5mm]
   \hs{2}0\hs{7}{\ds \f{-1}{\k-\ma}}\hs{0.25}(\gmn-{\ds \f{\kmu\knu}{\k}})
\hs{-2} 
\end{array}
\right) 
\eea
and
\vs{-2}
\bea
\label{d1-ab}
D_{ab}^{(1)}(k) = 
{\ds \fr{k^4}}\left(
\begin{array}{r}
\hs{2}-\g
 \hs{9}ig\knu\hs{1}  \\[5mm]
   \hs{-1}-ig\kmu\hs{4}-\kmu\knu\hs{-1} 
\end{array}
\right) 
\eea
$D_{ab}^{(0)}$ is block-diagonal, $D_{ab}^{(1)}(k)$ is
hermitian and the $\xi$-dependence is made explicit.
The matrix $V_{bc}^{\,3}(k,k')$ of 3-point ($\chi,\,A$)--vertices 
({\em i.e.} $\ro\chi\chi,\,\ro\chi A$ and $\ro A A$) can be
easily read from the table in Appendix D:
\bea
\label{v3}
V_{bc}^{\,3}(k,k') 
= 2\,i\,\ups
\left(
\begin{array}{r}
\hs{-1}-k\cdot k'  
 \hs{7}  i\,g\,k_\si \hs{-1}\\[2mm]
   i\,g\,k'_\nu   \hs{7} 
g_{\nu\,\si}\g\hs{-1} 
\end{array}
\right)
\eea
It is straightforward to check that
\be
\label{no-xi}
D^{(1)}_{ab}(k)\cdot V_{bc}^{\,3}(k,k') 
=0
\ee
So the $\xi$-dependence in Eq.(\ref{si-ro}) drops out and it becomes
\bea
\label{good-si}
\Si\hs{0.25}'(p) =
 \f{i}{2}\int \hs{-1}{\ds \f{d\hs{0.25}^dk}{(2\pi)^d}}
\left\{
\f{\lam^2\ups^2}{(\k - \mh)((p+k)^2 - \mh)}\, + 
\right.\hs{35}\non\\[4mm]
\left.\mb + \f{4\hs{0.25}\g\ma}{(\k - \ma)((p+k)^2 - \ma)}
\left(d - \f{\k}{\ma} - \f{(p+k)^2}{\ma} + \f{[k\cdot(p+k)]^2}{m_A^4}\right)
\right\}
\eea
where $d$ is the space-time dimension. 
The $p^2$-term of Eq.(\ref{good-si}) gives $Z_\ro(\ups)$. 
In essentially the same fashion, 
the $\xi$-dependence in $Z_1(\ups)$ also drops out.
It is straightforward to check that $Z_1(\ups)=Z_3(\ups)=Z_4(\ups)$.
Finally, to one loop we have
\bea
\label{Z-ro}
Z_\ro [\ro] & = & 1+ {\ds \f{\hbar}{16\pi^2}}\left\{ 
{\ds \f{\lam^2 \ro^2}{12\,\mh}} + 
                3\g\ln{\ds\f{\ma}{\mu^2}}\right\}\\[4mm]
Z_1 [\ro] & = & 1+ {\ds \f{\hbar}{16\pi^2}}
\left\{3\g{\ds \f{\mh\ln{\ds \f{\mh}{\mu^2}}\,-\,
\ma\ln{\ds \f{\ma}{\mu^2}}}{\mh \,-\, \ma}} 
 \,+\, 
{\ds \f{\mh\,-\,5\hs{0.25}\ma}{2\hs{0.25}\ups^2}}\right\}
\eea
where $\mu^2$  is the renormalization scale.

Analogously, to evaluate $Z_A[\ro]$ and $Z_{gf}[\ro]$ we consider the
1PI two-point function $\Si_A^{\mu\nu}(p)$, extract $\gmn p^2$ and 
$p^\mu p^\nu$ terms and determine $Z_A[\ro]$ and $Z_{gf}[\ro]$ from them.
Again, $\xi$--dependence drops out and
\bea
\label{A-si}
\Si_A^{\mu\nu}\hs{0.25}'(p) =
 -\f{i}{2}\int \hs{-1}{\ds \f{d\hs{0.25}^dk}{(2\pi)^d}}
\f{8\g(\ma\gmn - \kmu\knu )}{(\k - \ma)((p+k)^2 - \mh)} 
\eea
(where the prime denotes omission of $p^\mu$--independent diagrams
as before). 

One readily calculates $Z_A$:
\bea
\label{Z-A}
Z_A [\ro] & = & 1+ {\ds \f{\hbar}{16\pi^2}}\,\g\!
\left\{
\fr{3}{\ds \f{\ma\ln{\ds \f{\ma}{\mu^2}}-\mh\ln{\ds \f{\mh}{\mu^2}}}
{\ma \,-\, \mh}} 
 \,-\,{\ds \f{7m_A^4 + 4\ma\mh + m_H^4}{3(\ma \,-\, \mh)^2}} 
\right.\non\\
&&\left.\mb +\,{\ds \f{4m_A^4\mh\ln{\ds \f{\ma}{\mh}}}{(\ma \,-\, \mh)^3}}
\right\}
\eea

It is also straightforward to extract $Z_{gf}[\ro]$ from Eq.(\ref{A-si}).
It is obvious that after $\G\,[\,\phi, A_\mu ]$ is calculated in the 
polar basis, it can be re-expressed in {\em any} basis we want.

	Actually, the above $\xi$-independence is a general
feature of the complete $\G\,[\,\phi, A_\mu ]$ in perturbation expansion.
First, let us examine the reason why the $\xi$-dependence in the above 
calculations drops out. 
As seen in Eq.(\ref{no-xi}), the product of the $\xi$-dependent part 
$D_{ab}^{(1)}(k)$ of the ($\chi,\,A$)-propagator  with the 
vertex matrix $V_{bc}^{\,3}(k,k')$ (\ref{v3}) vanishes.
As the matrix $V_{bc}^{\,4}(k,k')$ of 4-point ($\chi,\,A$)--vertices
is proportional to $V_{bc}^{\,3}(k,k')$:
\vs{-1}
\be
\label{v3-v4}
V_{bc}^{\,4}(k,k') = \fr{\ups}\,V_{bc}^{\,3}(k,k')\;,
\vs{-1}\ee
we also have
\be
\label{no-xi-4}
D^{(1)}_{ab}(k)\cdot V_{bc}^{\,4}(k,k') = 0
\ee

The only $\xi$-dependence in the Feynman rules is in the 
$\xi\,D^{(1)}_{ab}$. Every propagator $iD_{ab}$ (\ref{polar-prop}) 
in an arbitrary 
n-point multi-loop diagram contracts with a vertex matrix, either
$V_{bc}^{\,3}$ or $V_{bc}^{\,4}$.
Applying (\ref{no-xi}) and/or (\ref{no-xi-4}) to {\em any} $n$-point
multi-loop diagram one can readily show its $\xi$-independence.
This means the full effective action, to {\em all} orders in perturbation 
expansion including higher derivative terms, is passively gauge-invariant. 
By construction, 
$\G\,[\,\phi, A_\mu]$ is also actively gauge-invariant. 
To conclude, the complete effective action $\G\,[\,\phi, A_\mu]$
evaluated by perturbative expansion in the polar gauge is explicitly 
gauge-invariant. 
	In the polar gauge, the Goldstone boson $\chi$ is always massless,
at or away from the minimum of the effective potential. So the Higgs 
mechanism can take place whenever $\ups$ is non-zero.

\section{Unitary Gauge}
\label{section-unitary}
With the gauge invariance condition (\ref{ZZZ}), independent
calculations of $Z_1[\ro]$, $Z_3[\ro]$ and $Z_4[\ro]$ are clearly redundant.
To avoid this redundancy, let us introduce the combination
\be
      B_\mu = A_\mu - \fr{g}\dm\chi
\ee
which brings us to the unitary gauge Lagrangian :

\bea
\cl & = &  \fr{2}\,(\dm\ro)^2 + 
\fr{2}\,\g B^2 \ro^2 - \fr{4}\,\widetilde{F}_{\mu\nu}\widetilde{F}^{\mu\nu} - 
V[\ro] +\, \ro\,\bar{c}\,c
\eea
In this case, a single source term for the combination $B_\mu$ can 
be introduced in the generating functional, instead of separate 
sources for $A_\mu$ and $\chi$.
In the unitary gauge, both $B_\mu$ and $\ro$ are
invariant under the gauge transformation (\ref{g-tr}),
so the source terms for $B_\mu$ and $\ro$ are gauge invariant.
Also, it is well-known that
only physical degrees of freedom have quantum fluctuations,
and no gauge-fixing is necessary in this case. So the
resulting $\G[\phi, A_\mu]$ is by definition gauge-invariant.

Consistency requires that the contributions in the unitary gauge and in the
 polar gauge must be identical. Before we demonstrate this, let us first
 consider the one-loop case.
The one-loop contribution to the effective potential in the 
unitary gauge has been calculated long ago \cite{doja,wein}: 
\bea
\label{EP-unit}
V_1(\ups) &=&- \f{i}{2} \int \hs{-1}{\ds \f{d\hs{0.25}^4k}{(2\pi)^4}}
\left\{\ln\left[\,- \g\,(\k - \ma)\hs{0.25}^3\,(\k - \mh)\right]\,-\,
2\ln\ups\right\}\non
\eea
So $V[\ro]$ in the unitary gauge
 agrees with $V[\ro]$ in the 
polar gauge (\ref{polact}).

 Evaluation of $\Si(p)$ is also straightforward (as before, 
$p_\mu$-independent terms are omitted):
\bea
\label{si-unit}
\Si\hs{0.25}'(p)&=& \f{i}{2}\int \hs{-1}{\ds \f{d\hs{0.25}^4k}{(2\pi)^4}}
\left\{D_H(k)(-i\lam\ups)D_H(p+k)(-i\lam\ups)\, + \right.\\[2mm]
&&\left.\mb + \,D_{\mu\nu}^U(k)\,(2\,i\hs{0.25}\g\ups\,g{\raisebox{-0.60ex}
{${\scriptstyle \nu\ro}$}})\,
D_{\ro\si}^U(p+k)\,(2\,i\hs{0.25}\g\ups\,g{\raisebox{-0.60ex}
{${\scriptstyle \si\mu}$}})
\right\}\non
\eea
where 
$D_H$ is the Higgs propagator and $D_{\mu\nu}^U$ is the gauge boson propagator
(\ref{d-u}) in the unitary gauge.
It is easy to check that Eq.(\ref{si-unit}) is exactly Eq.(\ref{good-si}), 
{\em i.e.}
the second term in Eq.(\ref{si-unit}) reproduces the $\xi$-independent
contributions from diagrams $(b,\hs{0.25}c,\hs{0.25}d)$ in Eq.(\ref{si-ro}). 
So 
 $Z_\ro[\ro]$ in the unitary gauge agrees with 
that obtained in the polar gauge.
The $\lam^2$ term in $Z_\ro[\ro]$ (\ref{Z-ro}) agrees with that 
given in Ref.\cite{iim}.
 As expected, $Z_1[\ro]$ is also the same in both gauges.
For the evaluation of $Z_A[\ro]$ one readily obtains exactly the same 
integrand as
(\ref{A-si}), thus $Z_A[\ro]$ is the same in both gauges as well.

It is straightforward to show the equivalence between the polar gauge
and the unitary gauge to any order of perturbation expansion.
This must be the case since both gauges yield the physical
effective action which is unique.
As in the one-loop case, they give the same integrands inside the momentum 
integrals in any multi-loop Feynman diagram. Here we shall give the 
key identity and briefly sketch how this equivalence appears in general.
The $\rho$-propagator and its part of the vertices are the same in 
the two gauges, so we only have to worry about the $(\chi,A)$-propagator 
and its vertices. Defining the unitary gauge vector-boson propagator 
matrix as
\bea
\label{d-uni}
\D_{ab}^{U}(k) = 
\left(
\begin{array}{r}
\hs{0}0
 \hs{11.5}0\hs{4}  \\[5mm]
   \hs{0}0\hs{6}D_{\mu\nu}^{U}(k)
\hs{-1} 
\end{array}
\right) ,
\eea
it is easy to check the following relation between the polar and the unitary 
gauges (we use notations introduced in the previous section):
\bea
\label{pol-uni}
V_{ab}\,(k_1,-k_2)\hs{0.75}D_{bc}(k_2)\hs{0.75}V_{cd}\,(k_2,-k_3)
\hs{1}=\hs{1}V_{ab}\,(k_1,\,0)\hs{0.75}\D_{bc}^U(k_2)\hs{0.75}
V_{cd}\,(0,\,-k_3)
 \eea
where every $V$ is either $V^{\,3}$ given in Eq.(\ref{v3}) or 
$V^{\,4}$ given in Eq.(\ref{v3-v4}).

Consider an arbitrary multi-loop Feynman diagram in the polar gauge. 
Suppose, inside this diagram, it has an internal loop of $(\chi,A)$-lines only
(as in Eq.(\ref{si-ro})).
Using Eqs.(\ref{d-uni},\,\ref{pol-uni}) and the cyclic property of the trace 
one can easily see that the following transformation can be done with the 
integrand:
\bea
\label{pol-uni2}
 {\rm Tr}\left[\,V\,(k_1,-k_2)\hs{0.75}D(k_2)
\hs{1}\ldots\hs{1}D(k_n)\hs{0.75}
V\,(k_n,-k_1)\hs{0.75}D(k_1)\,\right] = \\[2mm]
={\rm Tr}\left[\,V\,(0,0)\hs{0.75}\D^U(k_2)
\hs{1}\ldots\hs{1}\D^U(k_n)\hs{0.75}
V\,(0,0)\hs{0.75}\D^U(k_1)\,\right] = \hs{5.5}\non\\[2mm]
={\rm Tr}\left[\,V^U\hs{0.75}D^U(k_2)
\hs{1}\ldots\hs{1} D^U(k_n)\hs{0.75}V^U\hs{0.75}D^U(k_1)\,\right]\hs{26}\non
 \eea
where every $V^U$ is either $\ro A A$ or $\ro\ro A A$ vertex (same for
both gauges) and the trace in the last line is taken over the space-time
indices only.
 So every closed loop of $(\chi,A)$-lines in the polar gauge  
can be transformed into closed loop of $A$-lines in the unitary gauge. 

After rewriting every closed loop of this type in the unitary gauge, 
the only remaining $(\chi,A)$-propagators left are the ones that lead to 
external $(\chi,A)$-fields. In these cases,
we have ``open chains'' of $(\chi,A)$-lines. Each of them can be transformed as
\bea
\label{pol-uni3}
 V_{ab}\,(k_1,-k_2)\hs{0.75}D_{bc}(k_2)
\hs{1}\ldots\hs{1}D_{lm}(k_{n-1})\hs{0.75}
V_{mn}\,(k_{n-1},-k_n) = \\[2mm]
=V_{ab}\,(k_1,0)\hs{0.75}\D^U_{bc}(k_2)
\hs{1}\ldots\hs{1}\D^U_{lm}(k_{n-1})\hs{0.75}
V_{mn}\,(0,-k_n)\hs{8}\non 
 \eea
Recall that indices $a,b,\ldots$ label $\chi$ and $A$ fields together.
Once we specify that $a$ and $n$ take the values corresponding to 
external $A_\mu$ field only, we can easily see that
\bea
\label{pol-uni4}
V_{\mu b}\,(k_1,0)\hs{0.75}\D^U_{bc}(k_2)
\hs{1}\ldots\hs{1}\D^U_{lm}(k_{n-1})\hs{0.75}V_{m\nu}\,(0,-k_n)=\\[2mm]
=V^U_{\mu \si}\hs{0.75}D^U_{\si \ro}(k_2)
\hs{1}\ldots\hs{1}D^U_{\eta \lam}(k_{n-1})\hs{0.75}V^U_{\lam \nu}\hs{16}\non
 \eea
Thus, polar gauge and unitary gauge calculations yield the same integrands in
multi-loop calculations.

 	It is interesting to compare these two gauges.
Gauge-fixing is necessary in the polar gauge; this makes the calculation 
somewhat more complicated, but it has the advantage
that we can explicitly see how the $\xi$ parameter
disappears in the determination of $\G[\,\phi, A_\mu ]$.
In the unitary gauge calculations, there is no
gauge-fixing terms to worry about. However, even then, actively 
gauge-non-invariant terms (such as two-derivative terms involving the
gauge fields) will appear in the 
perturbation expansion. So we have to use the procedure outlined in
the introduction to obtain the gauge-invariant $\G[\,\phi, A_\mu ]$.

\section{Physical Effective Potential}
\label{section-physEP}
	 As we have seen, the kinetic terms in the effective action no
longer maintain their original canonical forms, {\em i.e.}, $Z_i$ are not 
unity. 
To define the physical effective potential that measures the
energy density in the universe, a field redefinition is
necessary. The Higgs field in the effective potential should be redefined
so that its kinetic term recovers its original canonical form. The 
effective potential in terms of this new dressed field should be the
physical effective potential.

	To illustrate this point, let us
consider for the moment the Lagrangian of a point
particle in classical mechanics.
Here the potential $U(x)$ measures the potential energy of
the particle. 
Now, let us consider $x=f(y)$, where $f(y)$ is
a well-behaving function. The action can be written as

\be
{\cal S} = \int\hs{-1.5} dt\, \left[\, \f{\;\dot{x}^2}{2} - U(x) \right] =
\int\hs{-1.5} dt\, \left[\,Z(y)\,\f{\;\dot{y}^2}{2} - V(y)\right]
\ee
where $Z(y)=f'(y)^2$. Of course, the theory is the same in either
coordinate, although $x$-coordinate is clearly the
natural and most physical choice.
In the $y$-coordinate, we need to know $V(y)$, as well as
the function $f(y)$ or, equivalently, $Z(y)$.
Starting in the $y$-coordinate,
the potential $U(x)$ cannot be determined from $V(y)$ alone - knowledge
of $Z(y)$ is also neccessary.
Now let us go back to the Higgs theory case where the situation
is completely analogous.
A new (dressed) field $\si(x)$ as a functional of $\ro(x)$
 can be introduced,  such that $Z_\si [\si] = 1$.
In the small momenta regime, $U[\si]$ determined by $V[\ro]$ and $Z_\ro[\ro]$
is the physical effective potential.
One may use it to calculate
physical masses, critical temperature and other physical quantities.

Let us introduce the dressed field $\si(x)$

\be
\si [\ro] = \int^\ro Z_\ro [\ro']^{\fr{2}}\: d\ro'
\ee

Now, the physical effective potential is the effective potential
expressed in terms of this dressed field $\si(x)$:
\bea
\label{phys-ep}
U[\si] &=& \, V[\ro\,(\si)] = \non\\[1mm]
&=& {\ds \f{m^2\si^2}{2}}\,+\,{\ds \f{\lam\si^4}{4!}}
\,+
{\ds \f{\hbar}{64\,\pi^2}}\left\{m_H^4(\,\ln{\ds \f{\mh}{\mu^2}} - \f{3}{2})\,+
3\,m_A^4(\,\ln{\ds \f{\ma}{\mu^2}} -\f{5}{6})\,-\right.\non\\[2mm]
&&\mb \left.
\,-\, 6\mg\ma(\,\ln{\ds \f{\ma}{\mu^2}} -2 ) \,+\, 
\mg\hs{0.25}{\ds \f{\lam\,\si}{3}}\left(\hs{-0.5}
a\,{\rm Arctanh}\hs{0.25}{\ds \f{\si}{\raisebox{0.6ex}{$a$}}}\,-\si
\right)
\hs{-1}\right\}
\eea
where 
\be
\ma = \g\si^2\,,\hs{5}\mh = m^2 + {\ds \f{\lam\si^2}{2}}\,,\hs{5}
\mg = m^2 + {\ds \f{\lam\si^2}{6}}\hs{3}{\rm and}\hs{3} a^2 = -\,{\ds \f{2m^2}
{\lam}}
\ee
For the spontaneous symmetry breaking case, $m^2$ is negative.

As an illustration, let us see how 
$U[\si]$ is related to the physical Higgs mass. Let the vacuum expectation 
values be $\langle\ro\rangle = \ups$ and $\langle\si\rangle = \om$.
Let us first assume that the higher derivative terms in the effective action
can be ignored. Then  
\be
\left.\left.m_H^2 = U''(\om) \right|\raisebox{-2ex}
{$ \om = \om_{min}$} = V''(\ups)\cdot Z^{-1}(\ups)\right|\raisebox{-2ex}
{$\ups = \ups_{min}$} 
\ee
where $\om_{min}$ and $\ups_{min}$ are determined by 
\bea
U'(\om_{min}) = 0 \hs{5}{\rm and}\hs{5}V'(\ups_{min}) = 0 
\non\eea

If the Higgs mass is not small, we must include the effects of the higher 
derivative terms in $\G[\phi, A_\mu]$. 
Recall that the 1PI Higgs two-point function gives
\be
\Sigma(p^2) = V''(\ups) - p^2 Z_\ro(\ups) - Y(p^2,\,\ups)
\ee
where $Y(p^2,\,\ups) \sim {\cal O}(p^4)$ and stands for the relevant higher
derivative terms, obtainable from Eq.(\ref{good-si}). 
In the polar gauge, $Y(p^2,\,\ups)$ is gauge-invariant
and finite at one-loop.
Looking for the Higgs mass as a pole 
in the propagator in the momentum space, we have
\be
\label{pole}
\left.Z_\ro(\ups)\,p^2 \,+\, Y(p^2, \ups) \,- \,V''(\ups)\right|\raisebox{-2ex}
{$ \ups = \ups_{min}
\,,\;p^2 = m_H^2$} = 0
\ee 
or
\be
\label{mass}
\mh = U''(\om_{min}) - Y(\mh, \ups_{min})\cdot Z^{-1}(\ups_{min})
\ee
To compare this result with that from the standard perturbation theory, let
us introduce the following notations:
\bea
 m_H^2 =  m_{H,\,0}^2 + \hbar\,m_{H,\,1}^2 +  
\ldots & &
\hs{10}\ups_{min} =  \ups_0 + \hbar\,\ups_1 + \ldots\non\\
Z_\ro = 1 + \hbar\,Z_{\ro,\,1} + \ldots \hs{8.75}& & 
\hs{10}\om_{min} =\om_0 + \hbar\,\om_1 + \ldots\non\\
V = V_0 + \hbar\,V_1 + \ldots \hs{10}&&\non   
\eea
Recall that 
the minimum of $V_0$ is at $\ups_0$ ($\om_0$ = $\ups_0$ since 
$Z_{\ro,\,0} = 1$) 
{\em i.e.} $V'_0(\ups_0) = 0$.
For $m^2 < 0$, we have 
\bea
\ups_0^2 = - {\ds \f{6\,m^2}{\lam}}\hs{10} {\rm and}\hs{10} 
m_{H,\,0}^2 = V''_0 (\ups_0)= -2m^2\non
\eea
To order $\hbar$, the minima of $V(\ups)$ and $U(\om)$ are
 shifted by $\ups_1$ and $\om_1$ correspondingly, with
\bea
\ups_1 = - {\ds \f{V'_1(\ups_0)}{V''_0(\ups_0)}}     
 \hs{10}{\rm and}\hs{10} \om_1 = \ups_1 + \fr{2}\int^{\ups_0}\hs{-2} 
Z_1(\ups ')\: d\ups '\non
\eea
and we have
\be
m_{H,\,1}^2 = V''_1(\ups_0) - m_{H,\,0}^2Z_1(\ups_0) - 
Y_1(m_{H,\,0}^2,\,\ups_0) - 
{\ds \f{V'_1(\ups_0)\,V'''_0(\ups_0)}{V''_0(\ups_0)}}
\ee
or
\be
\label{mass-dia}
m_{H,\,1}^2 = i\left[
\bpic(72,36)(-4,13.5)
\put(32,16){\circle{32}}
\put(16,16){\line(-1,0){16}}
\put(48,16){\line(1,0){16}}
\put(21,5){\line(1,1){22}}
\put(24.5,2.5){\line(1,1){21}}
\put(28.5,0.5){\line(1,1){18.5}}
\put(34.5,0.5){\line(1,1){13}}
\put(18.5,8.5){\line(1,1){21}}
\put(16.5,12.5){\line(1,1){19}}
\put(16.5,18.5){\line(1,1){13}}
\epic
+
\bpic(60,36)(12,4)
\put(32,16){\circle{32}}
\put(32,0){\line(0,-1){16}}
\put(16,-16){\line(1,0){32}}
\put(21,5){\line(1,1){22}}
\put(24.5,2.5){\line(1,1){21}}
\put(28.5,0.5){\line(1,1){18.5}}
\put(34.5,0.5){\line(1,1){13}}
\put(18.5,8.5){\line(1,1){21}}
\put(16.5,12.5){\line(1,1){19}}
\put(16.5,18.5){\line(1,1){13}}
\epic
\right]
\raisebox{-5ex}{${\textstyle \ups^2 = \ups_0^2\,,\;p^2 = m_{H,\,0}^2}$}
\ee
The last term corresponds to the second diagram  
:  $V'''_0(\ups_0)$ is $3$-vertex, $ - V''_0(\ups_0)$ is
 the inverse scalar propagator at zero momentum and $V'_1(\ups_0)$
is clearly the tadpole itself. This is simply the result derived in 
Ref.\cite{apcar}, where the gauge invariance of $m_{H,\,1}^2 $
is clearly shown.

Let us go back to the massless ({\em i.e.} $m^2 = 0$) 
scalar QED case. From Eq.(\ref{phys-ep}), 
we easily 
obtain the one-loop (physical) effective potential calculated in
the polar basis,
\bea
\label{popot}
U[\si] = {\ds \f{\lam\si^4}{4!}}\,+
{\ds \f{\hbar}{64\,\pi^2}}\left\{
\fr{4}\hs{0.25}\lam^2 + 3\hs{0.25}\g - \lam\hs{0.25}\g
\right\}\ln{\ds \f{\si^2}{\mu^2}} 
\eea
which is obviously $\xi$--independent. In the ($\hi,\go$)--basis in the 
covariant gauge, $Z_2[\phi]=0$ in the limit $m^2 = 0\,$, and
\bea
Z_\rho[\phi] = Z_1[\phi] = 1 \,+\, {\ds \f{\hbar}{16\,\pi^2}}\,(3-\xi)\hs{0.25}\g
\ln{\ds \f{\phi^2}{\mu^2}}
\eea
and, from Ref.\cite{cowe},
\bea
	V\raisebox{-0.8ex}{$\hs{-1}{\scriptstyle Cartesian}$}
[\phi]= {\ds \f{\lam\phi^4}{4!}}\,+
\,{\ds \f{\hbar}{64\,\pi^2}}\left\{
\f{5}{18}\hs{0.25}\lam^2 + 3\hs{0.25}\g - 
\fr{3}\hs{0.25}\xi\hs{0.25}\lam\hs{0.25}\g
\right\}\ln{\ds \f{\phi^2}{\mu^2}}
\eea
the resulting physical effective potential is also explicitly 
$\xi$--independent,
\bea
\label{car-ep}
	U\raisebox{-0.8ex}{$\hs{-1}{\scriptstyle Cartesian}$}[\si]=
{\ds \f{\lam\si^4}{4!}}\,+
\,{\ds \f{\hbar}{64\,\pi^2}}\left\{
\f{5}{18}\hs{0.25}\lam^2 + 3\hs{0.25}\g - \lam\hs{0.25}\g
\right\}\ln{\ds \f{\si^2}{\mu^2}}
\eea
so we expect all physically measurable quantities to be 
$\xi$-independent when calculated in this basis. However, the one-loop 
(physical) effective potential
calculated in the ($\hi,\go$)--basis, Eq.(\ref{car-ep}), differs from 
that calculated in the polar basis, Eq.(\ref{popot}).
Which is correct? Our reasoning clearly implies that the polar gauge 
calculation, Eq.(\ref{popot}), is correct. 
First, there is no symmetry associated with the massless limit.
When we move away from the massless limit, 
the $\xi$-dependence reappears in the ($\hi,\go$) basis result,
$U$\raisebox{-0.8ex}{$\hs{-1}{\scriptstyle Cartesian}$}[$\si$].
This can be explicitly checked, using the one-loop results for 
$Z_\rho[\phi]$ and $V[\phi]$ in the covariant gauge.
In contrast, Eq.(\ref{popot}) is always $\xi$-independent. 
Again, we can trace 
the difference (the $\hs{0.5}\lam^2/36\hs{1}$ term) 
between the two expressions to the non-zero Goldstone 
mass $\mg$ in the ($\hi,\go$)--basis. As we have argued, this 
$\mg$ is an artifact. 

\section{$SU(2)$ Higgs Theories}
\label{section-nonabel}
The above observations generalize to the non-Abelian cases. To be specific,
let us consider the $SU(2)$ gauge symmetry case with a $SU(2)$ doublet 
Higgs field, given by the Lagrangian (we denote $T^a A^{a\,\mu} \,= \,A^{\mu}$,
$T^a F_{\mu\nu}^a = F_{\mu\nu}$)
\bea
\label{non-ab}
{\cl}  =  -\fr{2}{\rm Tr}\left[F_{\mu\nu}F^{\mu\nu}\right]+\,\fr{2}\,
(D_\mu \phi)\raisebox{1ex}{\dag}(D^\mu \phi) - V[\phi]\hs{40}\\[2mm]
{\rm where}\hs{10}D^\mu = \dmm - igA^\mu\,,\hs{10}
F_{\mu\nu} = \dm A_\nu - \dn A_\mu - ig [A_\mu\,,\, A_\nu]
\non\eea
As usual, $T^a$'s are generators of the gauge group.
 We parametrize the scalar field as
\be
\phi(x) = U^{-1}(x)\ro(x) \hs{10}{\rm where}
\hs{10}U(x) = {\rm exp}\left\{-iT^c \chi^c(x)
\right\}
\ee
The unitary matrix field $U(x)$ contains would-be-Goldstone part of the Higgs 
sector and the column vector $\ro(x)$
contains remaining scalar fields.

 The effective action can be written as 
\bea
\label{ea-nab}
\G[\, \phi, A^a_\mu] 
&=& \!\!\int\! d\,^4x\left\{- V[\ro^2] \,+\, 
\fr{2}\,Z_\ro [\ro^2]\,\dm\ro^T\dmm\ro\,-\,\fr{2}\,Z_A[\ro^2]\,
{\rm Tr}\left[F_{\mu\nu}F^{\mu\nu}\right]\,+\,\right.\non 
\\[2mm]
&&\left.\mb+\, 
\fr{2}\,Z_1[\ro^2]\g\,\ro^T\left(U A^\mu U^{-1} 
- \f{i}{g}\left[\dm U\right]U^{-1}\right)^2\hs{-2}\ro\,
+ \ldots \right\}
\eea
where $\ro^2 = \ro^T\hs{-1}\ro$.
Each term in (\ref{ea-nab}) is actively gauge-invariant under the gauge 
transformation
\bea
\label{nab-gtr}
\ro(x) \lra \ro(x) \hs{20} U^{-1}(x) \lra W(x)U^{-1}(x)
\hs{15}\\[2mm]
A^\mu(x)\lra W(x)A^\mu(x)W^{-1}(x) - \f{i}{g}\left[
\dm W(x)\right]W^{-1}(x)\hs{15}\non
\eea
At the tree-level, $Z_A\,=\,Z_1\,=\,Z_\rho\,=\,1$. 
The easiest way to determine their
higher order contributions is to evaluate them in 
the unitary gauge, where we introduce 
\bea
\label{B-nab}
B^\mu(x)\,=\,U(x)A^\mu(x)U^{-1}(x) - \f{i}{g}\left[
\dm U(x)\right]U^{-1}(x) 
\eea
For example, the one-loop contribution to the effective potential is
\bea
V[\ro]= {\ds \f{m^2 \ro^2}{2} }+{\ds \f{\lam\,\ro^4}{4!}}+
{\ds \f{\hbar}{64\,\pi^2}}\left\{ m_H^4(\,\ln{\ds \f{\mh}{\mu^2}} - \f{3}{2}) +
9\,m_A^4(\,\ln{\ds \f{\ma}{\mu^2}} -\f{5}{6}) \right\}\\[3mm]
Z_\ro [\ro] = 1+  {\ds \f{\hbar}{16\,\pi^2}}
\left\{{\ds \f{\lam^2 \ro^2}{12\,\mh}} +
                {\ds \f{9\,\g}{4}}\ln{\ds\f{\ma}{\mu^2}}\right\}\hs{51}
\eea
where 
\be
\mh = m^2 + {\ds \f{\lam\,\ro^2}{2}}\,,\hs{10}\ma = {\ds \f{\g\ro^2}{4}}
\ee

\section{Finite Temperature}
\label{section-hot}	
	By now, it should be obvious that the perturbative contributions
to all orders to the
finite temperature effective potential $V^\beta [\ro]$ are also 
gauge-invariant. The $V^\beta[\ro]$ is obtained when we convert the
four-momentum integration in Eq.(\ref{EP-polar}) 
to a three-momentum integration and 
a sum over the Matsubara modes (in the imaginary time
formalism) \cite{fint,fint2},
\be
\int\hs{-1}{\ds \f{d\hs{0.25}^4k}{(2\pi)^4}} \lra 
\sum_\om \fr{\beta}\int \hs{-1}{\ds \f{d\hs{0.25}^3k}{(2\pi)^3}}\,,
\hs{20}k_0 \ra \om = 2\pi i\,nT
\ee
Up to irrelevant constant, the integrand in Eq.(\ref{EP-polar}) 
is $\xi$-independent.
This means that the $V^\beta [\ro]$ is gauge invariant, and
is exactly the same as that in the 
unitary gauge \cite{fint2}.
Clearly, the same gauge invariance argument applies to the various finite 
temperature $Z_i^\beta [\ro]$. Of course, the presence of the heat bath
four-velocity introduces additional invariants in the finite temperature
effective action \cite{weldon}.
The critical temperature (in the high temperature regime)
determined from the unitary gauge was shown to agree with
that determined from other gauges \cite{ueda,kelly}.

\section{Concluding Remarks}
\label{section-concl}
         It is an elementary fact that a problem with rotational symmetry 
should be tackled using polar coordinates, not Cartesian coordinates.
All Higgs theories have rotational symmetries, so any perturbation 
expansion should be carried out in the appropriate polar variable basis.

	What is wrong with the Cartesian basis? The problem 
stems from the  unphysical degrees of freedom with
gauge-dependent masses which are always present in the Cartesian basis
and thus bring  gauge-dependence into the effective action.
These gauge-dependent masses turn to zero when
the would-be Goldstone bosons mass vanishes. So, if we carry out the 
perturbation expansion around the minimum of the effective potential,
where the would-be Goldstone bosons are massless, the gauge 
dependence drops out. This explains why usual gauges in 
Cartesian basis are acceptable choices for the perturbation expansion 
around a stable point of the theory.

	The would-be Goldstone
bosons are always massless in the polar basis, no matter whether we sit
at the minimum of the potential or not (and we have to move away from the 
minimum in order to determine effective action).
Physically, the gauge bosons become
massive whenever the 
vacuum expectation value for the scalar field is non-zero.
For the Higgs mechanism to work, 
this requires the would-be Goldstone bosons to be massless, 
 at the minimum or
{\em away} from it.
Thus the polar basis should be used for the effective action calculations,
 while any basis is valid for the perturbative
calculations around the minimum of the potential. 

   Following the above observations, the procedure to obtain the
physical $\G[\phi,\,A_\mu]$ in perturbation expansion becomes obvious,
for both Abelian and non-Abelian Higgs theories, at zero or finite
temperature. First one calculates all terms up to a given loop and up to a
given number of derivatives; this gives $\G_{cov}[\phi,\,A_\mu]$.
Next one sets all the ghost background fields to zero. Then one removes all
terms that are not actively gauge-invariant. In particular, this removes the
terms coming from the gauge-fixing, such as the $Z_{gf}[\phi]$ term.
By definition, the resulting $\G[\phi,\,A_\mu]$ is actively gauge-invariant.

        When we apply the above procedure to the zeroth order, {\em i.e.},
the gauge-fixed tree-level effective action, we trivially recover the
original classical action. (This is equivalent to the BRST approach.
Recall that the gauge-fixed tree-level effective action is BRST invariant.
Now the nilpotency of the BRST operator allows us to separate
the BRST-closed part $\G[\phi,\,A_\mu]$ from the BRST-exact part (
the gauge-fixing terms and the ghosts).)
When we apply the above procedure in the general case,
the resulting physical $\G[\phi,\,A_\mu]$ obtained in the
covariant gauge in the polar basis is gauge invariant to all orders in
the perturbation expansion. This is explicitly demonstrated in the
one-loop approximation for $\G[\phi,\,A_\mu]$, which agrees with that
obtained in the unitary gauge.
As is clear from the above discussion, the resulting $\G[\phi,\,A_\mu]$ 
obtained in the covariant gauge in the Cartesian basis is
passively gauge-dependent and hence unphysical.

$\G\,[\,\phi, A_\mu ]$ obtained from one-loop calculations in the 
$(\hi , \go)$ basis is
explicitly $\xi$-dependent. We have shown that some physics obtained 
from the minima of $\G\,[\,\phi, A_\mu ]$ will also be $\xi$-dependent. 
What can we do about this problem ? To remove the explicit $\xi$-dependence,
we have to consider higher-loop contributions. However, from the 
expressions of the one-loop contributions, the only possibility that 
the explicit $\xi$-dependence can be removed is if all-loop contributions 
are taken into account. This is a daunting project.
Since the shape/geometry of the classical potential is given, the relation
between the Goldstone modes in the polar and the Cartesian bases is clear.
Then one must be able to relate the results in the polar basis to that in
the Cartesian basis; or equivalently, the results in the covariant gauge 
can be converted to
that in the unitary gauge. In fact, the mathematical formalism for doing
this, namely, the DeWitt-Vilkovisky approach, have been developed \cite{DV}.
Using this formalism, it has been explicitly shown 
(at least in the massless case) how the one-loop effective potential in the
covariant gauge can be converted to that in the unitary gauge \cite{kuns}.
It will be interesting to check if this approach is applicable in general.

	Even in the pure complex scalar theory, the effective potentials
obtained from the Cartesian and the polar bases are different. Again, the 
difference comes from the non-zero Goldstone boson mass in the Cartesian 
basis. Our argument implies 
that for theories with only 
global symmetries, the appropriate polar bases give reliable results. 

\section*{Acknowledgement}
\hs{10} We thank Mark Alford, Brian Greene, Zurab
Kakushadze, Tomislav Prokopec and Tung-Mow Yan for discussions. 
This work
is supported by a grant from the National Science Foundation.

\newpage
\section*{Appendix A}

\hs{5}Here we review \cite{doja} the details of the calculation of 
$V_{R_\xi,\,1}(\ups, u, \xi)$.
The quadratic (matter) part of the shifted and gauge-fixed 
Lagrangian (\ref {aga}) is :
\be 
\hs{-80} iD^{-1}(\ups, u, \xi) = 
\ee

\[ \left(
\begin{array}{l}
\hs{10}(\k - \mg)(\del{ij} - \eti\,\etj) + 
 \hs{30}  i\,g\,\ep{ii'}(\ups_{i'}-u_{i'})\,\knu \hs{5}\non\\

+ (\k - \mh)\,\eti\,\etj - M^2 ( \del{ij} - \vepi \vepj)
    \non\\
    \non\\
  \hs{15} -i\,g\,\ep{jj'}(\ups_{j'}-u_{j'})\,\kmu  \hs{20} 
- (\k-\ma)\left(\gmn-{\ds \f{\kmu\knu}{\k}}\right)- \non\\
\hs{75} - {\ds \fr{\xi} }\,(\k - \xi \ma)\,{\ds \f{\kmu\knu}{\k}} \non
\end{array}
\right) \]
\vs{2}\\
\mb\hs{10}where
\bea
 \ups^2 = \ups_1^2 + \ups_2^2\;,\hs{7}
\eti = {\ds \f{\ups_{i}}{\sqrt{\ups^2}}}\;, \hs{7}
 u^2 = u_1^2 + u_2^2\;,\hs{7} 
\vepi = {\ds \f{u_{i}}{\sqrt{u^2}}}\hs{5}
\\[2mm]
\mh = m^2 +  \f{\lam\ups^2}{2}\;,\hs{5}
\mg = m^2 +  \f{\lam\ups^2}{6}\;,\hs{5}
\ma = \g\ups^2\,, \hs{5} 
M^2 = \xi\,\g u^2 \non
\eea

Using the property 
\be
\det
\left|
\begin{array}{cc}
A & B\\
C & D
\end{array}
\right| 
= \det|A|\det|D-CA^{-1}B|\:,\hs{10} (\det|A|\neq 0)\:,
\ee

the evaluation of $\det|\,iD^{-1}(\ups, u, \xi)\,|$ is straightforward :
\be
\det|\,iD^{-1}(\ups, u, \xi)\,|=(\k - \ma)^3\,(\k - \mf)(\k - \ms)
(\k - \mt)
\ee

where $m_i$'s are defined by :
\bea
\label{alal}
\hs{-7}\mf + \ms +\mt = \mh + \mg + 2\,\xi\,\g\,u_i\,\ups_i\hs{41}\\[2mm]
\hs{-7} \mf\ms + \mf\mt + \ms\mt 
  = \mh\mg + \xi\ma\mg + 2\,\xi\,\g\mh\,u_i\,\ups_i +
\xi^2 g^4[u_i\,\ups_i]\,^2 \hs{-20}\non\\[2mm]
\hs{-7}\mf \ms \mt  =  \xi\ma\mh\mg +
\xi^2 g^4\mh [u_i\,\ups_i]^2 +
\xi^2 g^4\mg\,[\ep{ij}\ups_i\,u_j]\,^2 \hs{4}\non
\eea
The result is
$V_{R_\xi,\,1}(\ups, u, \xi)$,
with all
bare quantities replaced by their renormalized counterparts :
\bea
\label{V-rxi}
V_{R_\xi,\,1} (\ups, u, \xi)
= \fr{64\pi^2} \left\{  
	3m_A^4 \left(\ln{\ds \f{\ma}{\mu^2}} - {\ds \f{5}{6}}\right) +
        m_1^4 \left(\ln{\ds \f{\mf}{\mu^2}} - {\ds \f{3}{2}}\right)
\right. + \\[4mm]
\left.\mb + m_2^4 \left(\ln{\ds \f{\ms}{\mu^2}} - {\ds \f{3}{2}}\right) +
           m_3^4 \left(\ln{\ds \f{\mt}{\mu^2}} - {\ds \f{3}{2}}\right)
	  -2\,m_g^4 \left(\ln{\ds \f{m_g^2}{\mu^2}}-
{\ds \f{3}{2}}\right)\right\}\non
\eea
where the last term comes from the ghost contribution with 
$m_g^2 = \xi\g u_i \ups_i $.
\newpage
\section*{Appendix B}
The Feynman rules of the shifted Abelian Higgs model in the
covariant gauge, $(\phi_1, \phi_2)$ basis (\hs{0.25}
here we follow the
convention of Ref.\cite{kang}):

\begin{table}[tbh]
\centering
\begin{tabular}{|lcc|}
\multicolumn{3}{c}{\bf Propagators:}\\
\hline
Scalar:&
\bpic(52,30)
\put(0,0){$i$}
\put(44,0){$j$}
\put(7,3){\line(1,0){34}}
\epic
&$\mb\hs{3}i\left[{\ds \f{\k - \xi\ma}{D(\k)}}(\del{ij} - \eti\,\etj)\hs{4} 
+\hs{5} 
{\ds \fr{\k - \mh}}\eti\,\etj\hs{3}\right]$\\[5mm]
Gauge:&
\bpic(48,30)
\put(0,0){$\mu$}
\put(43,0){$\nu$}
\multiput(11,3)(8,0){4}{\oval(4,4)[t]}
\multiput(15,3)(8,0){4}{\oval(4,4)[b]}
\epic
&$-i\left[{\ds \fr{\k-\ma}}(\gmn-{\ds \f{\kmu\knu}{\k}})\hs{3} + \hs{1}
{\ds \f{\xi(\k-\mg)}{D(\k)}\f{\kmu\knu}{\k}}\right]$\\[5mm]
mixed:&
\bpic(52,24)
\put(0,0){$\mu$}
\put(46,0){$i$}
\multiput(11,3)(8,0){3}{\oval(4,4)[t]}
\multiput(15,3)(8,0){2}{\oval(4,4)[b]}
\multiput(35,3)(-0.1,0.1){30}{\line(1,0){0.1}}
\multiput(35,3)(-0.1,-0.1){30}{\line(1,0){0.1}}
\put(29,3){\line(1,0){16}}
\put(29,9){$k$}
\epic
&$-{\ds \f{\xi\,g}{D(\k)}}\,\kmu\,\ep{ij}\,\ups_{j}$\\
&&\\
\hline
\end{tabular}
\end{table}
\begin{table}[tbh]
\centering
\begin{tabular}{|cccc|}
\multicolumn{4}{c}{\bf Vertices:}\\
\hline
\bpic(40,48)(-8,0)
\put(16,16){\line(-1,-1){16}}
\put(16,16){\line(1,-1){16}}
\put(16,16){\line(0,1){21}}
\put(16,16){\circle*{3}}
\put(-8,-2){$j$}
\put(34,-2){$k$}
\put(18,36){$i$}
\epic
 & \raisebox{12pt}{$-i\,{\ds \f{\lam}{3}}\,(\del{ij}\ups_{k} + 
\del{jk}\ups_{j} + \del{ki}\ups_{j})$} &
\bpic(40,40)(-8,0)
\multiput(12,14)(-4,-4){4}{\oval(4,4)[br]}
\multiput(16,14)(4,-4){4}{\oval(4,4)[tr]}
\multiput(16,14)(-4,-4){4}{\oval(4,4)[tl]}
\multiput(20,14)(4,-4){4}{\oval(4,4)[bl]}
\put(16,16){\line(0,1){21}}
\put(16,16){\circle*{3}}
\put(-9,-2){$\mu$}
\put(34,-2){$\nu$}
\put(18,36){$i$}
\epic
&\raisebox{12pt}{$2\,i\g\ups_{i}\,\gmn$} \\[1mm]
\bpic(40,40)(-8,0)
\put(0,0){\line(1,1){32}}
\put(0,32){\line(1,-1){32}}
\put(16,16){\circle*{3}}
\put(-8,-2){$j$}
\put(34,-2){$l$}
\put(-7,24){$i$}
\put(34,24){$k$}
\epic
 & \raisebox{12pt}{$-i\,{\ds \f{\lam}{3}}\,(\del{ij}\del{kl} + 
\del{ik}\del{jl} + \del{il}\del{jk})$} &
\bpic(40,40)(-8,0)
\multiput(12,14)(-4,-4){4}{\oval(4,4)[br]}
\multiput(16,14)(4,-4){4}{\oval(4,4)[tr]}
\multiput(16,14)(-4,-4){4}{\oval(4,4)[tl]}
\multiput(20,14)(4,-4){4}{\oval(4,4)[bl]}
\put(16,16){\line(1,1){16}}
\put(16,16){\line(-1,1){16}}
\put(-9,-2){$\mu$}
\put(34,-2){$\nu$}
\put(-9,24){$i$}
\put(34,24){$j$}
\put(16,16){\circle*{3}}
\epic
&\raisebox{12pt}{$2\,i\g\del{ij}\,\gmn$}\\
\multicolumn{4}{|c|}{
\bpic(48,50)
\put(16,16){\line(1,-1){20}}
\multiput(16,14)(-4,-4){5}{\oval(4,4)[tl]}
\multiput(12,14)(-4,-4){5}{\oval(4,4)[br]}
\put(-13,-6){$\mu$}
\put(16,16){\line(0,1){25}}
\put(18,40){$i$}
\put(38,-6){$j$}
\put(16,16){\circle*{3}}
\multiput(16,26)(0.1,0.1){24}{\line(1,0){0.1}}
\multiput(16,26)(-0.1,0.1){24}{\line(1,0){0.1}}
\put(28,4){\line(-1,0){3}}
\put(28,4){\line(0,1){3}}
\put(30,8){$k_2$}
\put(2,28){$k_1$}
\epic
\raisebox{12pt}{$g\ep{ij}\,(k_{1}+k_2)\,^{\mu}$}}\\
&&&\\
\hline
\end{tabular}
\caption{Feynman rules in the covariant gauge}
\end{table}
where
\be
\label{m+m-}
\raisebox{12pt}{$D(\k) = k^4 - \k\mg + \xi\ma\mg = (\k - m_+^2)(\k -m_-^2)$}  
\ee
\newpage
\section*{Appendix C}

One-loop correction to the effective potential in the covariant gauge, $
(\hi, \go)$-basis, is simply given by the $u_1 = 0\,,\,u_2 = 0$ limit 
of Eq.(\ref{V-rxi}):
\bea
\label{V-cov}
V_{cov,\,1} (\ups)
= \fr{64\pi^2} \left\{  
   3m_A^4 \left(\ln{\ds \f{\ma}{\mu^2}} - {\ds \f{5}{6}}\right) +
   m_H^4  \left(\ln{\ds \f{\mh}{\mu^2}} - {\ds \f{3}{2}}\right) + 
\right.\\[4mm]
\left.\mb + m_+^4  \left(\ln{\ds \f{m_+^2}{\mu^2}} - {\ds \f{3}{2}}\right)  +
   m_-^4  \left(\ln{\ds \f{m_-^2}{\mu^2}} - {\ds \f{3}{2}}\right) \right\}
\non
\eea
where $m_+^2\,,\,m_-^2$ are defined by Eq.(\ref{m+m-})

To check the gauge symmetry condition (\ref{ZZZ}) we need to evaluate 
the following ten diagrams.
The $p^2$ term from diagrams $(a - c)$ contributes to $Z_1$,
the $p_\mu$ term from 
diagrams $(d - g)$ - to $Z_3$ and $p_\mu\, =\, 0$ term in
diagrams $(h - k)$ - to $Z_4 $.
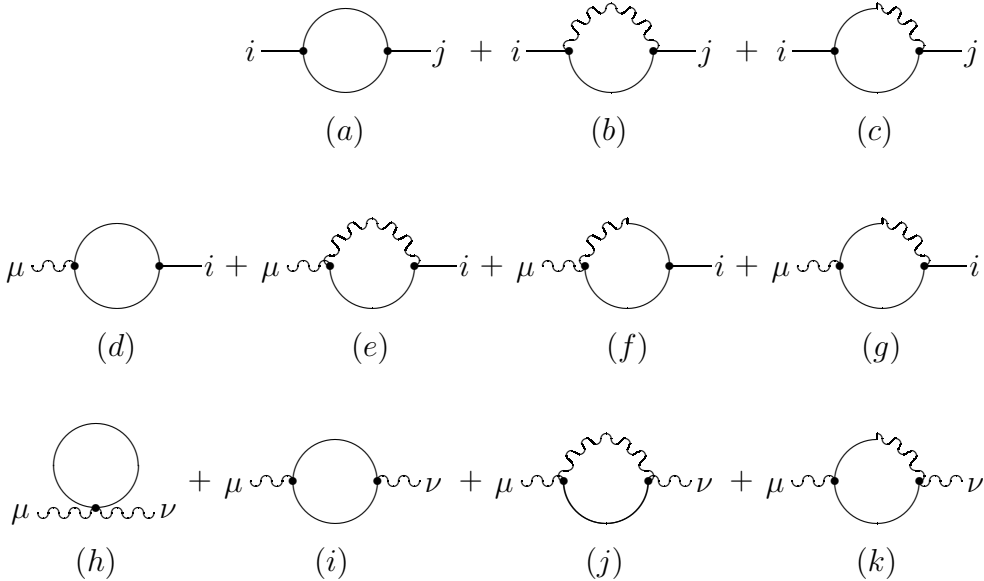
\bea
\mb\hs{-30}\bpic(76,40)(0,13)
\put(32,16){\circle{32}}
\put(16,16){\line(-1,0){16}}
\put(48,16){\line(1,0){16}}
\put(-6,12){$i$}
\put(65,12){$j$}
\put(16,16){\circle*{3}}
\put(48,16){\circle*{3}}
\put(24,-18){$(a)$}
\epic
+
\bpic(86,40)(-10,13)
\put(16,16){\circle*{3}}
\put(48,16){\circle*{3}}
\put(16,16){\line(-1,0){16}}
\put(48,16){\line(1,0){16}}
\put(32,16){\oval(32,32)[b]}
\put(-6,12){$i$}
\put(65,12){$j$}
\put(32,32){\oval(4,4)[t]}
\put(36,32){\oval(4,4)[bl]}
\put(28,32){\oval(4,4)[br]}
\multiput(36,30)(0.2,0.1){4}{\line(1,0){0.1}}
\multiput(28,30)(-0.2,0.1){4}{\line(1,0){0.1}}
\multiput(36.8,30.4)(0.1,0.1){16}{\line(1,0){0.1}}
\multiput(27.2,30.4)(-0.1,0.1){16}{\line(1,0){0.1}}
\multiput(38.4,32.0)(0.2,0.1){4}{\line(1,0){0.1}}
\multiput(25.6,32.0)(-0.2,0.1){4}{\line(1,0){0.1}}
\put(39.2,30.4){\oval(4,4)[tr]}
\put(24.8,30.4){\oval(4,4)[tl]}
\multiput(41.2,30.4)(-0.1,-0.2){12}{\line(0,1){0.1}}
\multiput(22.8,30.4)(0.1,-0.2){12}{\line(0,1){0.1}}
\put(42.0,28.0){\oval(4,4)[bl]}
\put(22.0,28.0){\oval(4,4)[br]}
\multiput(42.0,26.0)(0.2,0.1){15}{\line(1,0){0.1}}
\multiput(22.0,26.0)(-0.2,0.1){15}{\line(1,0){0.1}}
\put(45.0,25.5){\oval(4,4)[tr]}
\put(19.0,25.5){\oval(4,4)[tl]}
\multiput(47.0,25.5)(-0.1,-0.2){4}{\line(0,1){0.1}}
\multiput(17.0,25.5)(0.1,-0.2){4}{\line(0,1){0.1}}
\multiput(46.6,24.7)(-0.1,-0.1){15}{\line(0,1){0.1}}
\multiput(17.4,24.7)(0.1,-0.1){15}{\line(0,1){0.1}}
\multiput(45.1,23.2)(-0.1,-0.2){7}{\line(0,1){0.1}}
\multiput(18.9,23.2)(0.1,-0.2){7}{\line(0,1){0.1}}
\put(46.5,22.0){\oval(4,4)[bl]}
\put(17.5,22.0){\oval(4,4)[br]}
\multiput(46.5,19.9)(0.1,0){15}{\line(1,0){0.1}}
\multiput(17.5,19.9)(-0.1,0){15}{\line(1,0){0.1}}
\put(48.0,18.0){\oval(4,4)[r]}
\put(16.0,18.0){\oval(4,4)[l]}
\put(24,-18){$(b)$}
\epic
+
\bpic(70,40)(-10,13)
\put(-6,12){$i$}
\put(65,12){$j$}
\put(16,16){\circle*{3}}
\put(48,16){\circle*{3}}
\put(16,16){\line(-1,0){16}}
\put(48,16){\line(1,0){16}}
\put(32,16){\oval(32,32)[b]}
\put(32,16){\oval(32,32)[tl]}
\multiput(32,32)(0,0.2){10}{\line(0,1){0.1}}
\put(32,32){\oval(4,4)[tr]}
\put(36,32){\oval(4,4)[bl]}
\multiput(36,30)(0.2,0.1){4}{\line(1,0){0.1}}
\multiput(36.8,30.4)(0.1,0.1){16}{\line(1,0){0.1}}
\multiput(38.4,32.0)(0.2,0.1){4}{\line(1,0){0.1}}
\put(39.2,30.4){\oval(4,4)[tr]}
\multiput(41.2,30.4)(-0.1,-0.2){12}{\line(0,1){0.1}}
\put(42.0,28.0){\oval(4,4)[bl]}
\multiput(42.0,26.0)(0.2,0.1){15}{\line(1,0){0.1}}
\put(45.0,25.5){\oval(4,4)[tr]}
\multiput(47.0,25.5)(-0.1,-0.2){4}{\line(0,1){0.1}}
\multiput(46.6,24.7)(-0.1,-0.1){15}{\line(0,1){0.1}}
\multiput(45.1,23.2)(-0.1,-0.2){7}{\line(0,1){0.1}}
\put(46.5,22.0){\oval(4,4)[bl]}
\multiput(46.5,19.9)(0.1,0){15}{\line(1,0){0.1}}
\put(48.0,18.0){\oval(4,4)[r]}
\put(24,-18){$(c)$}
\epic\non\\
\non\\
\bpic(82,60)(-12,13)
\put(65,12){$i$}
\multiput(14,16)(-8,0){2}{\oval(4,4)[t]}
\multiput(10,16)(-8,0){2}{\oval(4,4)[b]}
\put(-10,12){$\mu$}
\put(16,16){\circle*{3}}
\put(48,16){\circle*{3}}
\put(48,16){\line(1,0){16}}
\put(32,16){\circle{32}}
\put(24,-18){$(d)$}
\epic
+
\bpic(82,60)(-12,13)
\put(65,12){$i$}
\multiput(14,16)(-8,0){2}{\oval(4,4)[t]}
\multiput(10,16)(-8,0){2}{\oval(4,4)[b]}
\put(-10,12){$\mu$}
\put(16,16){\circle*{3}}
\put(48,16){\circle*{3}}
\put(48,16){\line(1,0){16}}
\put(32,16){\oval(32,32)[b]}
\put(32,32){\oval(4,4)[t]}
\put(36,32){\oval(4,4)[bl]}
\put(28,32){\oval(4,4)[br]}
\multiput(36,30)(0.2,0.1){4}{\line(1,0){0.1}}
\multiput(28,30)(-0.2,0.1){4}{\line(1,0){0.1}}
\multiput(36.8,30.4)(0.1,0.1){16}{\line(1,0){0.1}}
\multiput(27.2,30.4)(-0.1,0.1){16}{\line(1,0){0.1}}
\multiput(38.4,32.0)(0.2,0.1){4}{\line(1,0){0.1}}
\multiput(25.6,32.0)(-0.2,0.1){4}{\line(1,0){0.1}}
\put(39.2,30.4){\oval(4,4)[tr]}
\put(24.8,30.4){\oval(4,4)[tl]}
\multiput(41.2,30.4)(-0.1,-0.2){12}{\line(0,1){0.1}}
\multiput(22.8,30.4)(0.1,-0.2){12}{\line(0,1){0.1}}
\put(42.0,28.0){\oval(4,4)[bl]}
\put(22.0,28.0){\oval(4,4)[br]}
\multiput(42.0,26.0)(0.2,0.1){15}{\line(1,0){0.1}}
\multiput(22.0,26.0)(-0.2,0.1){15}{\line(1,0){0.1}}
\put(45.0,25.5){\oval(4,4)[tr]}
\put(19.0,25.5){\oval(4,4)[tl]}
\multiput(47.0,25.5)(-0.1,-0.2){4}{\line(0,1){0.1}}
\multiput(17.0,25.5)(0.1,-0.2){4}{\line(0,1){0.1}}
\multiput(46.6,24.7)(-0.1,-0.1){15}{\line(0,1){0.1}}
\multiput(17.4,24.7)(0.1,-0.1){15}{\line(0,1){0.1}}
\multiput(45.1,23.2)(-0.1,-0.2){7}{\line(0,1){0.1}}
\multiput(18.9,23.2)(0.1,-0.2){7}{\line(0,1){0.1}}
\put(46.5,22.0){\oval(4,4)[bl]}
\put(17.5,22.0){\oval(4,4)[br]}
\multiput(46.5,19.9)(0.1,0){15}{\line(1,0){0.1}}
\multiput(17.5,19.9)(-0.1,0){15}{\line(1,0){0.1}}
\put(48.0,18.0){\oval(4,4)[r]}
\put(16.0,18.0){\oval(4,4)[l]}
\put(24,-18){$(e)$}
\epic
+
\bpic(82,60)(-12,13)
\put(65,12){$i$}
\multiput(10,16)(-8,0){2}{\oval(4,4)[t]}
\multiput(14,16)(-8,0){2}{\oval(4,4)[b]}
\put(-10,12){$\mu$}
\put(16,16){\circle*{3}}
\put(48,16){\circle*{3}}
\put(48,16){\line(1,0){16}}
\put(32,16){\oval(32,32)[b]}
\put(32,16){\oval(32,32)[tr]}
\multiput(32,32)(0,0.2){10}{\line(0,1){0.1}}
\put(32,32){\oval(4,4)[tl]}
\put(28,32){\oval(4,4)[br]}
\multiput(28,30)(-0.2,0.1){4}{\line(1,0){0.1}}
\multiput(27.2,30.4)(-0.1,0.1){16}{\line(1,0){0.1}}
\multiput(25.6,32.0)(-0.2,0.1){4}{\line(1,0){0.1}}
\put(24.8,30.4){\oval(4,4)[tl]}
\multiput(22.8,30.4)(0.1,-0.2){12}{\line(0,1){0.1}}
\put(22.0,28.0){\oval(4,4)[br]}
\multiput(22.0,26.0)(-0.2,0.1){15}{\line(1,0){0.1}}
\put(19.0,25.5){\oval(4,4)[tl]}
\multiput(17.0,25.5)(0.1,-0.2){4}{\line(0,1){0.1}}
\multiput(17.4,24.7)(0.1,-0.1){15}{\line(0,1){0.1}}
\multiput(18.9,23.2)(0.1,-0.2){7}{\line(0,1){0.1}}
\put(17.5,22.0){\oval(4,4)[br]}
\multiput(17.5,19.9)(-0.1,0){15}{\line(1,0){0.1}}
\put(16.0,18.0){\oval(4,4)[l]}
\put(24,-18){$(f)$}
\epic
+
\bpic(70,60)(-12,13)
\put(65,12){$i$}
\put(16,16){\circle*{3}}
\put(48,16){\circle*{3}}
\multiput(14,16)(-8,0){2}{\oval(4,4)[t]}
\multiput(10,16)(-8,0){2}{\oval(4,4)[b]}
\put(-10,12){$\mu$}
\put(48,16){\line(1,0){16}}
\put(32,16){\oval(32,32)[b]}
\put(32,16){\oval(32,32)[tl]}
\multiput(32,32)(0,0.2){10}{\line(0,1){0.1}}
\put(32,32){\oval(4,4)[tr]}
\put(36,32){\oval(4,4)[bl]}
\multiput(36,30)(0.2,0.1){4}{\line(1,0){0.1}}
\multiput(36.8,30.4)(0.1,0.1){16}{\line(1,0){0.1}}
\multiput(38.4,32.0)(0.2,0.1){4}{\line(1,0){0.1}}
\put(39.2,30.4){\oval(4,4)[tr]}
\multiput(41.2,30.4)(-0.1,-0.2){12}{\line(0,1){0.1}}
\put(42.0,28.0){\oval(4,4)[bl]}
\multiput(42.0,26.0)(0.2,0.1){15}{\line(1,0){0.1}}
\put(45.0,25.5){\oval(4,4)[tr]}
\multiput(47.0,25.5)(-0.1,-0.2){4}{\line(0,1){0.1}}
\multiput(46.6,24.7)(-0.1,-0.1){15}{\line(0,1){0.1}}
\multiput(45.1,23.2)(-0.1,-0.2){7}{\line(0,1){0.1}}
\put(46.5,22.0){\oval(4,4)[bl]}
\multiput(46.5,19.9)(0.1,0){15}{\line(1,0){0.1}}
\put(48.0,18.0){\oval(4,4)[r]}
\put(24,-18){$(g)$}
\epic\non\\
\non\\
\bpic(76,60)(-12,13)
\put(0,2){$\mu$}
\put(56,2){$\nu$}
\put(32,6){\circle*{3}}
\multiput(16,4)(8,0){5}{\oval(4,4)[t]}
\multiput(12,4)(8,0){6}{\oval(4,4)[b]}
\put(32,22){\circle{32}}
\put(24,-18){$(h)$}
\epic
+
\bpic(88,60)(-12,13)
\put(-10,12){$\mu$}
\put(66,12){$\nu$}
\put(16,16){\circle*{3}}
\put(48,16){\circle*{3}}
\multiput(14,16)(-8,0){2}{\oval(4,4)[t]}
\multiput(10,16)(-8,0){2}{\oval(4,4)[b]}
\multiput(50,16)(8,0){2}{\oval(4,4)[t]}
\multiput(54,16)(8,0){2}{\oval(4,4)[b]}
\put(32,16){\circle{32}}
\put(24,-18){$(i)$}
\epic
+
\bpic(88,60)(-12,13)
\put(-10,12){$\mu$}
\put(66,12){$\nu$}
\put(16,16){\circle*{3}}
\put(48,16){\circle*{3}}
\multiput(14,16)(-8,0){2}{\oval(4,4)[t]}
\multiput(10,16)(-8,0){2}{\oval(4,4)[b]}
\multiput(50,16)(8,0){2}{\oval(4,4)[t]}
\multiput(54,16)(8,0){2}{\oval(4,4)[b]}
\put(32,16){\oval(32,32)[b]}
\put(32,16){\oval(32,32)[b]}
\put(32,32){\oval(4,4)[t]}
\put(36,32){\oval(4,4)[bl]}
\put(28,32){\oval(4,4)[br]}
\multiput(36,30)(0.2,0.1){4}{\line(1,0){0.1}}
\multiput(28,30)(-0.2,0.1){4}{\line(1,0){0.1}}
\multiput(36.8,30.4)(0.1,0.1){16}{\line(1,0){0.1}}
\multiput(27.2,30.4)(-0.1,0.1){16}{\line(1,0){0.1}}
\multiput(38.4,32.0)(0.2,0.1){4}{\line(1,0){0.1}}
\multiput(25.6,32.0)(-0.2,0.1){4}{\line(1,0){0.1}}
\put(39.2,30.4){\oval(4,4)[tr]}
\put(24.8,30.4){\oval(4,4)[tl]}
\multiput(41.2,30.4)(-0.1,-0.2){12}{\line(0,1){0.1}}
\multiput(22.8,30.4)(0.1,-0.2){12}{\line(0,1){0.1}}
\put(42.0,28.0){\oval(4,4)[bl]}
\put(22.0,28.0){\oval(4,4)[br]}
\multiput(42.0,26.0)(0.2,0.1){15}{\line(1,0){0.1}}
\multiput(22.0,26.0)(-0.2,0.1){15}{\line(1,0){0.1}}
\put(45.0,25.5){\oval(4,4)[tr]}
\put(19.0,25.5){\oval(4,4)[tl]}
\multiput(47.0,25.5)(-0.1,-0.2){4}{\line(0,1){0.1}}
\multiput(17.0,25.5)(0.1,-0.2){4}{\line(0,1){0.1}}
\multiput(46.6,24.7)(-0.1,-0.1){15}{\line(0,1){0.1}}
\multiput(17.4,24.7)(0.1,-0.1){15}{\line(0,1){0.1}}
\multiput(45.1,23.2)(-0.1,-0.2){7}{\line(0,1){0.1}}
\multiput(18.9,23.2)(0.1,-0.2){7}{\line(0,1){0.1}}
\put(46.5,22.0){\oval(4,4)[bl]}
\put(17.5,22.0){\oval(4,4)[br]}
\multiput(46.5,19.9)(0.1,0){15}{\line(1,0){0.1}}
\multiput(17.5,19.9)(-0.1,0){15}{\line(1,0){0.1}}
\put(48.0,18.0){\oval(4,4)[r]}
\put(16.0,18.0){\oval(4,4)[l]}
\put(24,-18){$(j)$}
\epic
+
\bpic(72,60)(-12,13)
\put(-10,12){$\mu$}
\put(66,12){$\nu$}
\put(16,16){\circle*{3}}
\put(48,16){\circle*{3}}
\multiput(14,16)(-8,0){2}{\oval(4,4)[t]}
\multiput(10,16)(-8,0){2}{\oval(4,4)[b]}
\multiput(50,16)(8,0){2}{\oval(4,4)[b]}
\multiput(54,16)(8,0){2}{\oval(4,4)[t]}
\put(32,16){\oval(32,32)[b]}
\put(32,16){\oval(32,32)[tl]}
\multiput(32,32)(0,0.2){10}{\line(0,1){0.1}}
\put(32,32){\oval(4,4)[tr]}
\put(36,32){\oval(4,4)[bl]}
\multiput(36,30)(0.2,0.1){4}{\line(1,0){0.1}}
\multiput(36.8,30.4)(0.1,0.1){16}{\line(1,0){0.1}}
\multiput(38.4,32.0)(0.2,0.1){4}{\line(1,0){0.1}}
\put(39.2,30.4){\oval(4,4)[tr]}
\multiput(41.2,30.4)(-0.1,-0.2){12}{\line(0,1){0.1}}
\put(42.0,28.0){\oval(4,4)[bl]}
\multiput(42.0,26.0)(0.2,0.1){15}{\line(1,0){0.1}}
\put(45.0,25.5){\oval(4,4)[tr]}
\multiput(47.0,25.5)(-0.1,-0.2){4}{\line(0,1){0.1}}
\multiput(46.6,24.7)(-0.1,-0.1){15}{\line(0,1){0.1}}
\multiput(45.1,23.2)(-0.1,-0.2){7}{\line(0,1){0.1}}
\put(46.5,22.0){\oval(4,4)[bl]}
\multiput(46.5,19.9)(0.1,0){15}{\line(1,0){0.1}}
\put(48.0,18.0){\oval(4,4)[r]}
\put(24,-18){$(k)$}
\epic\non\\[4mm]
\non
\eea
Contributions of these diagrams are summarized in the following table:
\begin{table}[bth]
\centering
\begin{tabular}{|l|}
\hline
\raisebox{-2ex}{$\hs{-0.5}a)\hs{3}{\ds \f{\lam^2\ups^2}{9}}
\left\{D^1 - \xi\ma C^1\right\} $}
\\[8mm]
$\hs{-0.5}b)\hs{3}
\g\!\left(\fr{2} - 3B_0^0(A,H)\right) + $\\[3mm]
$\mb\hs{12} + \xi\g\!\left\{D^0 - \mg C^0\! + \mg (2\mh - \xi\ma)C^1\! - 2\mh D^1\! -\fr{2}
\right\}\hs{-2} $ 
\\[4mm]
$\hs{-0.5}c)\hs{3} 
2 \cdot\xi\g {\ds \f{\lam\ups^2}{3}}
\left\{ \mh C^1 - C^0 \right\} $ 
\\[4mm] 
\hline
\raisebox{-2ex}{$\hs{-0.5}d)\hs{3}{\ds \f{\lam^2 \ups^2}{9}}\left\{D^1 - 
\xi\ma C^1\right\} + 
\xi\g {\ds \f{\lam \ups^2}{3}}D^1 $}
\\[8mm] 
$\hs{-0.5}e)\hs{3}
\g\left(\fr{2} - 3B_0^0(A,H)\right) + 
\xi\g\left\{ \mg(\mh - \xi\ma)C^1 - \mh D^1 - \fr{2} \right\}$ 
\\[4mm]
$\hs{-0.5}f)\hs{3}\xi\g {\ds \f{\lam \ups^2}{3}}
\left\{\mh C^1 - C^0 -D^1\right\} $
\\[4mm] 
$\hs{-0.5}g)\hs{3}\xi \g \left\{
D^0 + {\ds \f{\lam \ups^2}{3}}D^1 + \xi\ma\mg C^1 + \fr{2}\right\} $
\\[4mm]
\hline 
\raisebox{-2ex}{$\left.\begin{array}{l}\hs{-2}h\hs{-2}\\\hs{-2}i\hs{-2}
\end{array}\right\}\hs{2}{\ds \f{\lam^2 \ups^2}{9}}
\left\{D^1 - \xi\ma C^1\right\} + \xi\g\left\{2{\ds \f{\lam\ups^2}{3}}D^1
 + \xi\ma\mg C^1 \right\}$ }
\\[8mm] 
$\hs{-0.5}j)\hs{3} \g\left(\fr{2} - 3B_0^0(A,H)\right) + 
 \xi\g \left\{ \mg C^0 - D^0 - \fr{2}\right\}$ 
\\[4mm]
$\hs{-0.5}k)\hs{3}2 \cdot \xi \g \left\{D^0 + \fr{2}\right\} $ 
\\[4mm]
\hline
\end{tabular}
\caption{} 
\end{table}

Following notations were used for the integrals calculated in the
minimal substraction scheme  :
\bea
{\ds \f{i}{16\pi^2}}B_0(m_1, m_2, p^2)\, \dot{=}\,
\mu_0^{ \epsilon } \int {\ds \f{d\,^d k}{(2\pi)^d}}\,{\ds \fr{(k^2 - m^2_1)
((k + p)^2 - m^2_2)}}\\[4mm]
B_0(m_1, m_2, p^2) = B_0^0(m_1, m_2) + B_0^1(m_1, m_2)\cdot p^2 + O(p^4) + 
\ldots 
\eea
Shorthand $B(m_1, m_2) = B(1, 2)$ is used from now on.
\bea
B_0^0(1, 2) &=& {\ds \f{m_1^2\ln{\ds \f{m_1^2}{\mu^2}} - 
m_2^2\ln{\ds \f{m_2^2}{\mu^2}}}{m_2^2 - m_1^2}}\\[2mm]
B_0^1(1, 2) &=& \fr{2}{\ds \f{m_1^2 + m_2^2}{(m_1^2 - m_2^2)^2}} + {\ds
\f{m_1^2 m_2^2 \ln{\ds \f{m_1^2}{m_2^2}}}{(m_1^2 - m_2^2)^3}}\non
\eea
where 
\be
\ln\mu^2 \equiv {\ds \f{2}{\epsilon}} - \gamma + \ln4\pi\mu_0^2 \non
\ee
As a shorthand, we introduce
\be
C^i = {\ds \f{B^i_0 (H, +) - B^i_0 (H, -)}{m^2_+ - m^2_-}};\hs{10}
D^i = {\ds \f{m^2_+ B^i_0 (H, +) - m^2_- B^i_0 (H, -)}{m^2_+ - m^2_-}}
\ee
with $m_+$ and $m_-$ defined by Eq.(\ref{m+m-}).

Tensor integrals were reduced to scalar integrals and expressed in terms of 
$B_0^0$ and $B_0^1$ with the help of covariant decomposition and contractions.
An overall factor $\f{i}{16\pi^2}$ is implicit throughout.\vs{1}From the Table
 2, it is straightforward to check that
$Z_1 = Z_3 = Z_4$.

   $Z_\ro = Z_1 + Z_2$ is evaluated in a similar way, leading to the following 
cumbersome expression for the 1--loop correction 
(one can easily extract $Z_2$ from it since $Z_1$ is already known):
\bea
Z_{\ro,\hs{0.25}1}[\phi] \hs{-2}& = &\hs{-2} - 3\g\left({\ds \f{\mpl - \xi\ma}{\mpl - \mmi}}B_0^0(A,+)\,+
\,{\ds \f{\mmi - \xi\ma}{\mmi - \mpl}}B_0^0(A,-) - 1\right) + 
\non\\[4mm]
&+& \xi\g \left(B_0^0(+,-) + \fr{3}\right) \,+\, 
{\ds \f{\lam^2 \phi^2}{12\,\mh}} \,+\, {\ds \f{\lam^2 \phi^2}{108\,\mg}}+
\non\\[4mm]  
&+& {\ds \f{\ln{\ds \f{\mpl}{\mmi}}}{(\mpl - \mmi)^3}}\,\xi^2\ma\mg
\left(4\hs{0.25}\xi\g\ma - 2\g\mg + \f{\lam}{3}\hs{0.25}\ma
\right)+\non\\[4mm]
&+&3{\ds \f{\xi\g\ma}{\mpl - \mmi}}\left(
{\ds \f{\mpl{\ds \ln\f{\mmi}{\ma}}}{\ma - \mmi}} \,-\,
{\ds \f{\mmi{\ds \ln\f{\mpl}{\ma}}}{\ma - \mpl}}
\right)+\non\\[4mm]
&+&6{\ds \f{\xi\g\ma}{\mpl - \mmi}}\left(
{\ds \f{\mmi\ln\mmi - \ma\ln\ma}{\ma - \mmi}}\,-\,
{\ds \f{\mpl\ln\mpl - \ma\ln\ma}{\ma - \mpl}}+
\right)\non\\[4mm]
&+&{\ds \f{\xi\ma}{(\mpl - \mmi)^2}}\left(
\f{2}{3}\lam\hs{0.25}\xi\ma \,+\,\f{10}{3}\hs{0.25}\xi\hs{0.25}\g\mg \,+\, 
{\ds \f{\lam^2 \phi^2}{108}} \,-\, \f{4}{9}\lam\hs{0.25}\mg
\right)+\non\\[4mm]
&+&{\ds \f{B_0^1(+,-)}{(\mpl - \mmi)^2}}\hs{0.25}\xi^2 m_A^4\left(
\f{4}{3}\hs{0.25}\lam\hs{0.25}\mg \,-\, {\ds \f{\lam^2 \phi^2}{9}} \,-\, 
16\hs{0.25}\xi\hs{0.25}\g\mg 
\right)
\eea

Once again, we want to emphasize that $Z_2$ does not vanish.
\newpage
\section*{Appendix D}
Shifting $\ro \ra \ro + \ups$ in the Lagrangian (\ref{L-polar}) 
one gets the following Feynman rules in the polar gauge
(neglecting trivial ghost fields):
\begin{table}[tbh]
\centering
\begin{tabular}{|lcc|}
\multicolumn{3}{c}{\bf Propagators:}\\
\hline
$\ro$-field:&
\bpic(52,24)
\put(7,3){\line(1,0){34}}
\epic
&$\mb\hs{3}{\ds \f{i}{\k - \mh}}$\\
Gauge:&
\bpic(48,30)
\put(0,0){$\mu$}
\put(43,0){$\nu$}
\multiput(11,3)(8,0){4}{\oval(4,4)[t]}
\multiput(15,3)(8,0){4}{\oval(4,4)[b]}
\epic
&$-i\left[{\ds \fr{\k-\ma}}(\gmn-{\ds \f{\kmu\knu}{\k}})\hs{3} + \hs{1}
{\ds \f{\xi\kmu\knu}{k^4}}\right]$\\
$\chi$-field:&
\bpic(52,24)
\multiput(7,3)(6,0){6}{\line(1,0){2.5}}
\epic
&$i\,{\ds \f{\k - \xi\ma}{k^4 \ups^2}}$
\\
mixed:&
\bpic(52,24)
\put(0,0){$\mu$}
\multiput(11,3)(8,0){3}{\oval(4,4)[t]}
\multiput(15,3)(8,0){2}{\oval(4,4)[b]}
\multiput(45,3)(-0.2,0.2){15}{\line(1,0){0.1}}
\multiput(45,3)(-0.2,-0.2){15}{\line(1,0){0.1}}
\multiput(29.4,3)(6,0){4}{\line(1,0){2.8}}
\put(37,9){$k$}
\epic
&${\ds -\,\f{\xi g k_\mu}{k^4}}$\\
&&\\
\hline
\end{tabular}
\end{table}
\begin{table}[tbh]
\centering
\begin{tabular}{|cccccc|}
\multicolumn{6}{c}{\bf Vertices:}\\
\hline
\bpic(32,48)
\put(16,16){\line(-1,-1){16}}
\put(16,16){\line(1,-1){16}}
\put(16,16){\line(0,1){21}}
\put(16,16){\circle*{3}}
\epic
 & \raisebox{12pt}{$ -i\lam\ups $} &
\bpic(44,40)(-10,0)
\multiput(12,14)(-4,-4){4}{\oval(4,4)[br]}
\multiput(16,14)(4,-4){4}{\oval(4,4)[tr]}
\multiput(16,14)(-4,-4){4}{\oval(4,4)[tl]}
\multiput(20,14)(4,-4){4}{\oval(4,4)[bl]}
\put(16,16){\line(0,1){21}}
\put(16,16){\circle*{3}}
\put(-9,-2){$\mu$}
\put(34,-2){$\nu$}
\epic
&\raisebox{12pt}{$2\,i\g\ups\,\gmn$} &
\bpic(40,40)(-4,0)
\put(16,16){\line(0,1){21}}
\put(16,16){\circle*{3}}
\multiput(16,16)(0.1,-0.1){30}{\line(1,0){0.1}}
\multiput(21,11)(0.1,-0.1){20}{\line(1,0){0.1}}
\multiput(25,7)(0.1,-0.1){20}{\line(1,0){0.1}}
\multiput(29,3)(0.1,-0.1){20}{\line(1,0){0.1}}
\multiput(16,16)(-0.1,-0.1){30}{\line(1,0){0.1}}
\multiput(11,11)(-0.1,-0.1){20}{\line(1,0){0.1}}
\multiput(7,7)(-0.1,-0.1){20}{\line(1,0){0.1}}
\multiput(3,3)(-0.1,-0.1){20}{\line(1,0){0.1}}
\put(20.5,11.5){\line(1,0){4}}
\put(20.5,11.5){\line(0,-1){4}}
\put(11.5,11.5){\line(-1,0){4}}
\put(11.5,11.5){\line(0,-1){4}}
\put(27,8){$k_2$}
\put(-6,8){$k_1$}
\epic
&\raisebox{12pt}{$-2\,i\ups (k_1\cdot k_2)$}\\
&&&&&\\
\bpic(32,40)
\put(0,0){\line(1,1){32}}
\put(0,32){\line(1,-1){32}}
\put(16,16){\circle*{3}}
\epic
 & \raisebox{12pt}{$-i\lam $} &
\bpic(44,40)(-10,0)
\multiput(12,14)(-4,-4){4}{\oval(4,4)[br]}
\multiput(16,14)(4,-4){4}{\oval(4,4)[tr]}
\multiput(16,14)(-4,-4){4}{\oval(4,4)[tl]}
\multiput(20,14)(4,-4){4}{\oval(4,4)[bl]}
\put(16,16){\line(1,1){16}}
\put(16,16){\line(-1,1){16}}
\put(-9,-2){$\mu$}
\put(34,-2){$\nu$}
\put(16,16){\circle*{3}}
\epic
&\raisebox{12pt}{$2\,i\g\,\gmn $}&
\bpic(40,40)(-4,0)
\put(16,16){\line(1,1){16}}
\put(16,16){\line(-1,1){16}}
\put(16,16){\circle*{3}}
\multiput(16,16)(0.1,-0.1){30}{\line(1,0){0.1}}
\multiput(21,11)(0.1,-0.1){20}{\line(1,0){0.1}}
\multiput(25,7)(0.1,-0.1){20}{\line(1,0){0.1}}
\multiput(29,3)(0.1,-0.1){20}{\line(1,0){0.1}}
\multiput(16,16)(-0.1,-0.1){30}{\line(1,0){0.1}}
\multiput(11,11)(-0.1,-0.1){20}{\line(1,0){0.1}}
\multiput(7,7)(-0.1,-0.1){20}{\line(1,0){0.1}}
\multiput(3,3)(-0.1,-0.1){20}{\line(1,0){0.1}}
\put(20.5,11.5){\line(1,0){4}}
\put(20.5,11.5){\line(0,-1){4}}
\put(11.5,11.5){\line(-1,0){4}}
\put(11.5,11.5){\line(0,-1){4}}
\put(27,8){$k_2$}
\put(-6,8){$k_1$}
\epic
&\raisebox{12pt}{$-2\,i(k_1\cdot k_2)$}\\
&&&&&\\
\multicolumn{3}{|c}{\bpic(48,48)(-4,-4)
\put(16,16){\line(0,1){21}}
\put(16,16){\circle*{3}}
\multiput(16,14)(4,-4){4}{\oval(4,4)[tr]}
\multiput(20,14)(4,-4){4}{\oval(4,4)[bl]}
\put(34,-2){$\mu$}
\multiput(16,16)(-0.1,-0.1){30}{\line(1,0){0.1}}
\multiput(11,11)(-0.1,-0.1){20}{\line(1,0){0.1}}
\multiput(7,7)(-0.1,-0.1){20}{\line(1,0){0.1}}
\multiput(3,3)(-0.1,-0.1){20}{\line(1,0){0.1}}
\put(11.5,11.5){\line(-1,0){4}}
\put(11.5,11.5){\line(0,-1){4}}
\put(-3,8){$k$}
\epic
\raisebox{12pt}{$2g\ups k^\mu$}}
&\multicolumn{3}{c|}{\bpic(48,48)(-4,-4)
\put(16,16){\line(1,1){16}}
\put(16,16){\line(-1,1){16}}
\put(16,16){\circle*{3}}
\multiput(16,14)(4,-4){4}{\oval(4,4)[tr]}
\multiput(20,14)(4,-4){4}{\oval(4,4)[bl]}
\put(34,-2){$\mu$}
\multiput(16,16)(-0.1,-0.1){30}{\line(1,0){0.1}}
\multiput(11,11)(-0.1,-0.1){20}{\line(1,0){0.1}}
\multiput(7,7)(-0.1,-0.1){20}{\line(1,0){0.1}}
\multiput(3,3)(-0.1,-0.1){20}{\line(1,0){0.1}}
\put(11.5,11.5){\line(-1,0){4}}
\put(11.5,11.5){\line(0,-1){4}}
\put(-3,8){$k$}
\epic
\raisebox{12pt}{$2gk^\mu$}}\\
\hline
\end{tabular}
\caption{Feynman rules in the polar gauge}
\end{table}

In the unitary gauge the $\chi$-field is absent. So
 the Feynman rules in the unitary gauge 
can be obtained from the above tables if we delete all propagators
and vertices that involve $\chi$ and replace the gauge propagator with:
\be
\label{d-u}
D_{\mu\nu}^{U}(k) = -\,{\ds \f{i}{\k-\ma}}\,(\,\gmn-{\ds \f{\kmu\knu}{\ma}})  
\ee

\end{document}